\documentclass[onecolumn,preprintnumbers,superscriptaddress,amsmath,amssymb, bibnotes,nofootinbib]{revtex4}
\usepackage{amsthm}
\usepackage{amsmath}
\usepackage{amssymb}
\usepackage{graphicx}
\usepackage{url}
\usepackage{dsfont}
\usepackage{float}
\usepackage{bm}
\usepackage{ifthen}
\usepackage[usenames,dvipsnames]{color}
\usepackage{mathrsfs}
\usepackage[colorlinks=true,citecolor=blue,urlcolor=black]{hyperref}
\usepackage{float}
\usepackage{afterpage}
\usepackage{subfigure}
\usepackage{adjustbox}
\usepackage{times}
\usepackage{multirow}
\usepackage{tabularx}
\usepackage{booktabs}

\newcolumntype{L}[1]{>{\raggedright\arraybackslash}p{#1}}
\newcolumntype{C}[1]{>{\centering\arraybackslash}p{#1}}
\newcolumntype{R}[1]{>{\raggedleft\arraybackslash}p{#1}}

\begin{document}
	
%title
\title{Proof-of-principle experimental demonstration of twin-field type quantum key distribution}
\author{Xiaoqing Zhong}
\email{xzhong@physics.utoronto.ca}
\affiliation{Center for Quantum Information and Quantum Control, Dept. of Physics, University of Toronto, Toronto, Ontario, M5S 1A7, Canada}
\author{Jianyong Hu}
\email{jianyong\_hu@163.com}
\affiliation{Center for Quantum Information and Quantum Control, Dept. of Electrical \& Computer Engineering, University of Toronto, Toronto, Ontario, M5S 3G4, Canada}
\author{Marcos Curty}
\affiliation{EI Telecomunicaci\'on, Dept. of Signal Theory and Communications, University of Vigo, E-36310 Vigo, Spain}
\author{Li Qian}
\affiliation{Center for Quantum Information and Quantum Control, Dept. of Electrical \& Computer Engineering, University of Toronto, Toronto, Ontario, M5S 3G4, Canada}
\author{Hoi-Kwong Lo}
\affiliation{Center for Quantum Information and Quantum Control, Dept. of Physics, University of Toronto, Toronto, Ontario, M5S 1A7, Canada}
\affiliation{Center for Quantum Information and Quantum Control, Dept. of Electrical \& Computer Engineering, University of Toronto, Toronto, Ontario, M5S 3G4, Canada}

%\date{\today}
%%%%%%%%%%%%%%%%%%%%%%%%%%%%%%%%%%%%%%%%%%%%%%%%%%%%%%%%%%%%%%%%%%%%%%%
% Abstract
%%%%%%%%%%%%%%%%%%%%%%%%%%%%%%%%%%%%%%%%%%%%%%%%%%%%%%%%%%%%%%%%%%%%%%%
\begin{abstract}
	The twin-field (TF) quantum key distribution (QKD) protocol and its variants are highly attractive because they can beat the well-known rate-loss limit ({\it i.e.}, the PLOB bound) for QKD protocols without quantum repeaters. In this paper, we perform a proof-of-principle experimental demonstration of TF-QKD based on the protocol proposed by Curty et al.~\cite{tf-qkd_marcos} which removes from the original TF-QKD scheme the need for post-selection on the matching of a global phase, and can deliver nearly an order of magnitude higher secret key rate. Furthermore, we overcome the major difficulty in the practical implementation of TF-QKD, namely, the need to stabilize the phase of the quantum state over kilometers of fiber. A Sagnac loop structure is utilized to ensure excellent phase stability between the different parties. Using decoy states, we demonstrate secret-key generation rates that beat the PLOB bound when the channel loss is above 40 dB. \vspace{\baselineskip}
		
	
\end{abstract}

\maketitle
%%%%%%%%%%%%%%%%%%%%%%%%%%%%%%%%%%%%%%%%%%%%%%%%%%%%%%%%%%%%%%%%%%%%%%%%%%%%%%%%%%%%%%%
\textit{Introduction} - Quantum key distribution (QKD)~\cite{R1,R2,R3,R4,Ekert1991} makes it possible to distribute secret keys to remote users with information-theoretic security, which means that its security is independent of an attacker's computational power~\cite{R5,R6,R7,R8,R9,R10,R11,NJP11,Mayer_proof,secure}. Experimentally, QKD has been performed over 421 km of fiber~\cite{R13}, as well as over 1000 km of free space through satellite to ground links~\cite{R14,R15}. Towards the construction of a global quantum internet, performing long distance QKD is a crucial step~\cite{R16,R17,R18,R19}. However, there is a fundamental limit on the point-to-point secret key rate versus channel transmittance~\cite{plob,R21,R22} that can be achieved by two remote parties using QKD without an intermediate node. This limit, also called the PLOB bound~\cite{plob}, states that the secret key rate scales basically linearly with the channel transmittance.

To overcome the PLOB bound, besides using quantum repeaters~\cite{R23,R24,R25,R26}, it has been proposed to employ measurement-device-independent (MDI) QKD~\cite{R29} in combination with quantum memories~\cite{PRA89,NJP16} or in combination with quantum non-demolition measurements~\cite{R27} located at the untrusted intermediate node that is used in MDI-QKD. While promising, all these approaches are however far away from our current experimental capabilities. 
Remarkably, more recently it has been theoretically proven~\cite{R39,R40,R41,R42,tf-qkd_marcos,R44,finite_decoy_tfqkd} that variations of the twin-field (TF) QKD protocol proposed by Lucamarini et al.~\cite{tf-qkd_original} can beat the PLOB bound with the help of just one untrusted intermediate node (Charlie) performing a simple interferometric measurement. This shows that intercity QKD could be feasible with today’s technology without the need for quantum memories or quantum non-demolition measurements. In TF-QKD, two users (Alice and Bob) send two optical fields to produce a single-photon interference on a beam splitter located at Charlie. A successful result corresponds to Charlie observing a single-photon detection event, which measures the relative phase between the two optical beams. The fact that TF-QKD uses singles (i.e. single-photon detection events) results in a secret key rate that scales as the square-root of the channel transmittance instead of linearly. This is because only one photon (either from Alice or from Bob) has to arrive at Charlie. Importantly, since TF-QKD has a similar structure as MDI-QKD, in the sense that it uses an untrusted node to interfere Alice and Bob’s signals, all the advantages of MDI-QKD~\cite{R29,PRA89,NJP16,R27,R30,R31,R32,R33,R34}, such as its immunity to any possible attack against the measurement unit and its readiness for star networks, are retained by TF-QKD. In this regard, TF-QKD can be considered as a MDI-QKD scheme based on singles, rather than on coincidences.  

With the foundations of TF-QKD firmly established, it is now very important to demonstrate the viability of TF-QKD experimentally. One main drawback of the original TF-QKD protocol for practical implementations is that it requires long-distance subwavelength path-length phase stability, which is a new requirement in QKD and is much more demanding to achieve than two-photon interference as needed in standard MDI-QKD. Another drawback is that the original protocol requires to perform a post-selection step based on the matching of a global phase between Alice and Bob, which results in a reduction of the secret key rate by about an order of magnitude. To overcome these limitations, various variants of the original TF-QKD protocol have been very recently proposed and investigated~\cite{R39,R40,R41,R42,tf-qkd_marcos,R44,finite_decoy_tfqkd}. For example, ref.~\cite{R39} analyzes the security of a modified TF-QKD scheme by exploiting a quantum coin idea~\cite{R6,R10}. Ref.~\cite{R40}, on the other hand, introduces a phase-matching QKD which releases the requirement of active global phase randomization, but still needs a phase post-selection step. In Refs.~\cite{tf-qkd_marcos,R44,finite_decoy_tfqkd} the need for such post-selection step is removed. In summary, the security of some variants of TF-QKD have now been firmly established and their key rates beat the PLOB bound. So, it is now important to implement experimentally a TF-QKD protocol to demonstrate their practicality.

In this paper we perform a proof-of-principle experimental implementation of a TF-QKD protocol introduced by Curty et al.~\cite{tf-qkd_marcos}. This protocol does not need a post-selection step based on the matching of a global phase, and can provide a secret key rate which is about an order of magnitude higher than previous proposals~\cite{qcrypt2018}. The key idea is to use coherent states for key generation and photon number states as the complementary basis to prove security~\cite{NJP11}. The latter type of states can be simulated by means of phase-randomized coherent states in combination with the decoy-state method~\cite{R35,R36,R37,R38}. In our experiment, to stabilize the phase of the quantum states over kilometers of fiber, an auto-compensating set-up, which provides excellent phase stability, is built up. We remark that an auto-compensating set-up, also known as “plug-and-play” system, is widely used in QKD~\cite{plug-play,NJP4} and is the workhorse of a widely deployed commercial QKD system manufactured by the company ID Quantique. Security proofs for such “plug-and-play” QKD systems have been developed in~\cite{PRA77,NJP12}. More concretely, we use a Sagnac loop where an optical pulse (generated by a single laser in our experiment) travels through either a clockwise path or an anti-clockwise path and then the two paths interfere with each other. In one path, the pulse is modulated in phase by Alice whereas in the other path, the pulse is modulated in phase by Bob. To implement the decoy-state method, intensity modulators are employed to modulate the pulses leaving Bob and Alice’s stations. Our experimental results confirm an achievable secret key rate well above the PLOB bound, and constitute a crucial step towards demonstrating the practical experimental feasibility of TF-QKD.

\textit{Protocol and Experiment} - The TF-QKD protocol introduced in~\cite{tf-qkd_marcos} is composed of the following five steps.

Step 1: Each of Alice and Bob prepares a weak coherent state. Alice (Bob) chooses the $X$ basis with probability $P_X$ and the $Z$ basis with probability $P_Z=1-P_X$. If the $X$ basis is chosen, Alice (Bob) randomly prepares a coherent state $\left| \alpha\right\rangle_A$ ($\left| \alpha\right\rangle_B$) for the bit value $b_A=0$ ($b_B=0$) or $\left|-\alpha\right\rangle_A$ ($\left|-\alpha\right\rangle_B$) for the bit value $b_A=1$ ($b_B=1$). If the $Z$ basis is chosen, Alice (Bob) prepares a phase-randomized coherent state 
\begin{equation}\label{state}
\rho_A=\frac{1}{2\pi}\int_{0}^{2\pi}d\varphi_A\left|\beta_Ae^{i\varphi_A}\right\rangle_A\left\langle\beta_Ae^{i\varphi_A}\right|
\end{equation}
($\rho_B$ has the same expression as~(\ref{state}) with all subscripts changed to $B$). The value of the intensity $|\beta_A|^2$ ($|\beta_B|^2$) is chosen at random from a set $S=\left\lbrace \mu,\nu,\omega\right\rbrace $ containing say three possible intensities. 

Step 2: Alice and Bob send their states to the middle node Charlie through optical channels, each of them with transmittance $\sqrt{\eta}$.

Step 3: Charlie interferes the incoming states with a 50:50 beam splitter followed by two single-photon detectors, $D_0$ and $D_1$. He records the result, i.e. which detector clicks at each expected arrival time slot. 

Step 4: Once the quantum communication phase of the protocol has finished, Charlie announces all the results obtained, and Alice and Bob declare the bases used. 

Step 5: Based on the information announced, Alice and Bob estimate the bit and phase error rate and distill a secret key from those instances where they used the $X$ basis and Charlie declared one detection click. More precisely, whenever Charlie reports one click event in say $D_0$ ($D_1$) and both Alice and Bob choose the $X$ basis, $b_A$ and $b_B$ ($b_B\oplus1$) are regarded as their raw keys. 

\begin{figure}[b]
	{\includegraphics[width=12.9cm]{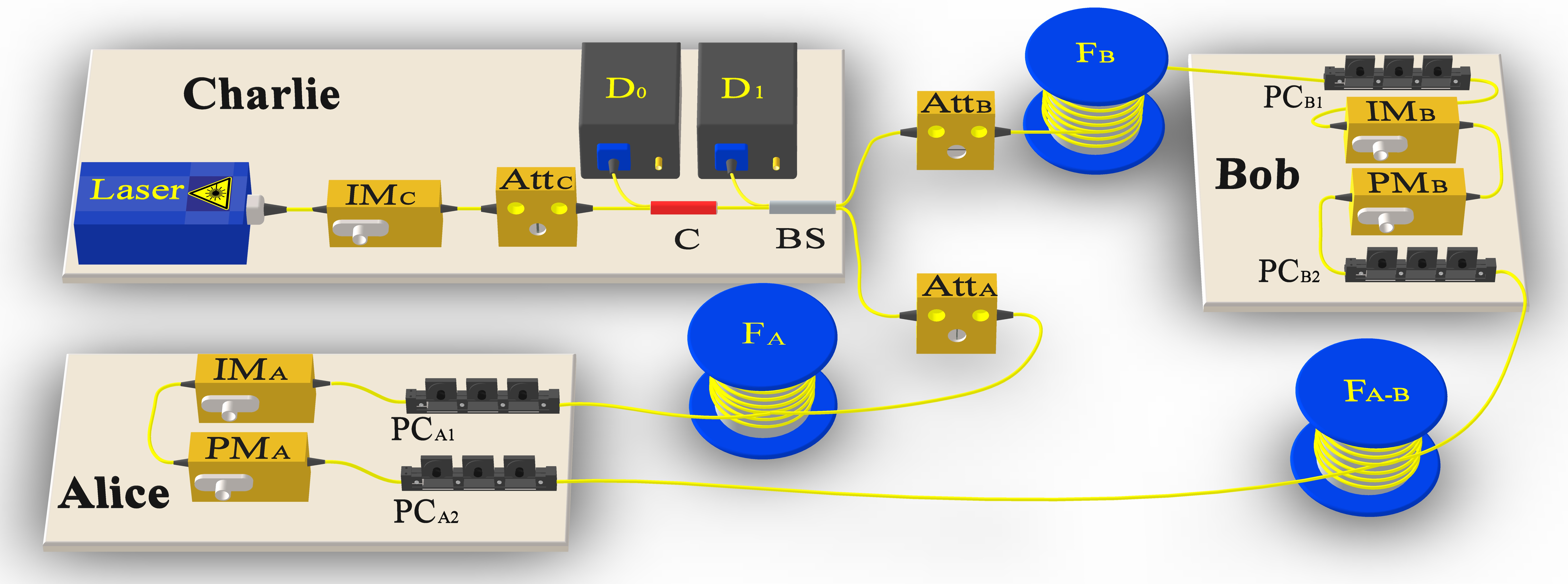}}
	\caption{Experimental set-up of twin-field quantum key distribution. The experiment starts by Charlie modulating the cw light from the laser to create coherent pulses through his intensity modulator (IM$_C$). Then he uses the optical attenuator (Att$_C$) to prepare weak coherent pulses. The pulses travel through a circulator (C) and enter the Sagnac loop by a 50:50 beam splitter (BS). The clockwise (counter-clockwise) traveling pulse goes through another optical attenuator Att$_B$ (Att$_A$) and, for some cases, a 5-km single mode fiber spool F$_B$ (F$_A$) before reaching at Bob’s (Alice’s) station. Bob (Alice) then lets the pulse pass through his (her) station with no modulation and sends it to Alice (Bob) through a 7-km fiber spool F$_{A-B}$. Once Alice (Bob) receives the pulse, she (he) uses her (his) phase modulator PM$_A$ (PM$_B$) and intensity modulator IM$_A$ (IM$_B$) to add a phase to the pulse and modulate the intensity of the pulse respectively, based on her (his) choice of the basis and her (his) bit value. Then Alice (Bob) sends the modulated coherent pulse back to Charlie’s BS through F$_A$ (F$_B$) and Att$_A$ (Att$_B$). Charlie records the interference result by using two single photon detectors $D_0$ and $D_1$, and publicly announces the outcomes.}
	\label{set-up}
\end{figure}

Fig.~\ref{set-up} shows the schematic diagram of our experimental set-up. It is a two-way QKD system consisting of a Sagnac interferometer. It is similar to the “plug-and-play” QKD system employed in~\cite{plug-play,NJP4}. The Sagnac arrangement is chosen to overcome the main challenges in implementing TF-QKD, namely, to share a phase reference between Alice and Bob and to achieve single-photon interference at Charlie, which requires phase stability. The common-path nature of the Sagnac loop automatically compensates for phase fluctuations of the two fields from Alice and Bob. In this set-up, the laser source is in Charlie's hands and is shared by Alice and Bob. This ensures that the three parties have the same phase reference. Charlie uses his intensity modulator (IM$_C$) and his optical attenuator (Att$_C$) to create weak coherent pulses from a cw DFB laser (PRO 800, wavelength 1552.6 nm). The IM$_C$ has an extinction ratio of $>$30 dB. The pulses have a FWHM width of 900 ps and a repetition rate of 10 MHz. They go through an optical circulator and then enter the Sagnac loop through a 50:50 fiber-based beam splitter. Clockwise and counter-clockwise traveling pulses each go through an optical variable attenuator (and a 5-km fiber spool F$_B$ or F$_A$ respectively in one case) before arriving at Alice’s or Bob's station. The clockwise (counter-clockwise) pulses go through Bob’s (Alice’s) station without being modulated. Therefore, no information is directly communicated between Alice and Bob. The clockwise (counter-clockwise) pulses then undergo a 7-km fiber spool F$_{A-B}$ and reach Alice's (Bob's) station. This is to ensure that Alice and Bob are physically separated with kilometers of fiber. Inside the station of Alice (Bob), there is a phase modulator PM$_A$ (PM$_B$) and an intensity modulator IM$_A$ (IM$_B$). If Alice (Bob) chooses the $X$ basis, she (he) uses her (his) PM$_A$ (PM$_B$) to randomly add a $0$ or $\pi$ phase to the pulse. If Alice (Bob) chooses the $Z$ basis, then she (he) uses her (his) PM$_A$ (PM$_B$) to add a random phase between $0$ and $2\pi$ to the pulse. The intensity modulator is applied to set the average number of photons per pulse to be, either $|\alpha|^2$ for the signal states in the $X$ basis, or one of the intensities in the set $S=\left\lbrace \mu,\nu,\omega\right\rbrace$ for the decoy states in the $Z$ basis. After the phase and intensity modulations, Alice (Bob) sends the pulses through a variable optical attenuator and, in some cases, through a 5-km spool of actual fiber F$_A$ (F$_B$), before reaching the beam splitter of Charlie. The loss between Alice (Bob) and Charlie is adjusted to simulate the loss due to the communication channel. To demonstrate the practicality of the scheme, we add the 5-km fiber spool F$_A$ (F$_B$) between Alice (Bob) and Charlie, in addition to the attenuator. The pulses coming from Alice and Bob interfere at Charlie’s beam splitter. One output of this beam splitter is directed to a single-photon detector (SPD) $D_0$ via the circulator, while the other output is followed directly by another SPD, $D_1$. The SPDs are commercial free-run avalanche photodiodes (ID220) with an efficiency of 11.7\% and a dark count rate of 750 Hz. The SPDs have a time jitter on the order of 200 ps, matching the optical pulse width. Charlie records each click event (within a 900 ps window where the detection is expected), and publicly announces the result. Afterward, Alice and Bob declare their bases choices and use the instances where they both selected the $X$ basis and Charlie announced a single-click event to distill a secure secret key.

Whether in Alice’s station or in Bob’s station, both clockwise and counter-clockwise traveling pulses pass through the IM and PM. It is crucial to ensure that Alice (Bob) only modulates the clockwise (counter-clockwise) traveling pulse. This is achieved by using appropriate fiber lengths, between Alice (Bob) and Charlie and between Alice and Bob, so that the two counter propagating pulses never overlap with each other at any modulator inside the Sagnac loop. Note that this is not a practical limitation, since in practice Alice and Bob can measure the fiber lengths in the link and add or remove small lengths of fiber within their own set-up. All modulators used in our set-up are driven and synchronized by a high-speed arbitrary waveform generator (AWG, Keysight M8195A). The delay times of the driving signal of Alice’s (Bob’s) IM$_A$ (IM$_B$) and PM$_A$ (PM$_B$), relative to the driving signal of Charlie’s IM$_C$, are well adjusted to ensure that Alice and Bob modulate the intended pulses. As in any practical system, there are unintended reflections and backscattering from the channel, causing unintended “clicks” in the detectors. Fortunately, these unintended signals do not arrive at the detectors at the same time as the signals from Alice and Bob. With precise synchronization, we can eliminate the unintended clicks by choosing the appropriate time windows for detection in Charlie's station. To further reduce the errors in the detection results, the fiber lengths within Charlie's station are adjusted to guarantee that the unintended signals due to reflections / backscattering do not overlap with the real signals at the detectors, i.e., do not fall inside the detection window. 

\begin{figure*} [t]
	\subfigure[]{\includegraphics[clip, trim=3cm 8cm 4cm 9cm, width=3.375in]{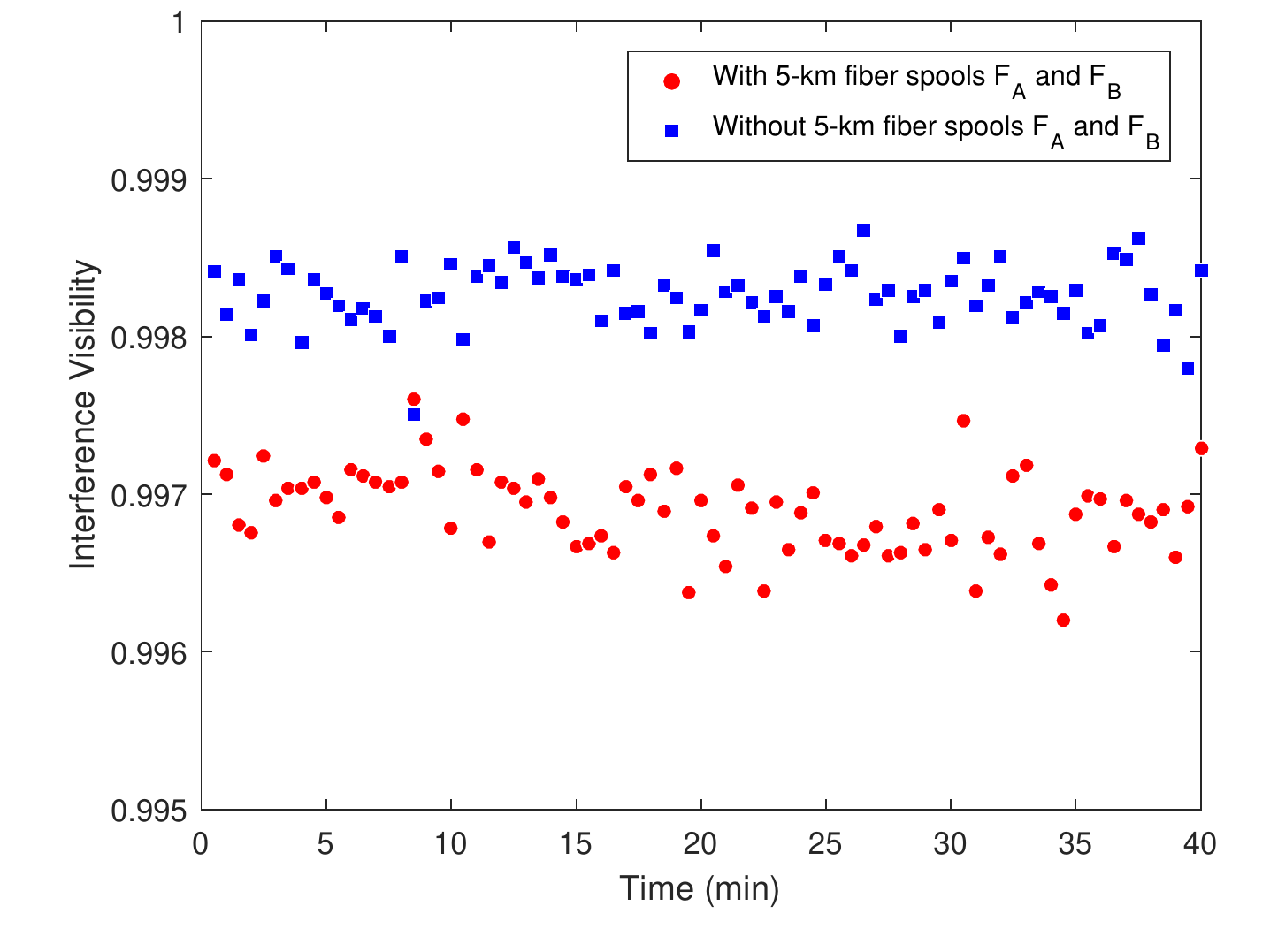}}
	\label{IV}	
	\subfigure[]{\includegraphics[clip, trim=3cm 8cm 4cm 9cm, width=3.375in]{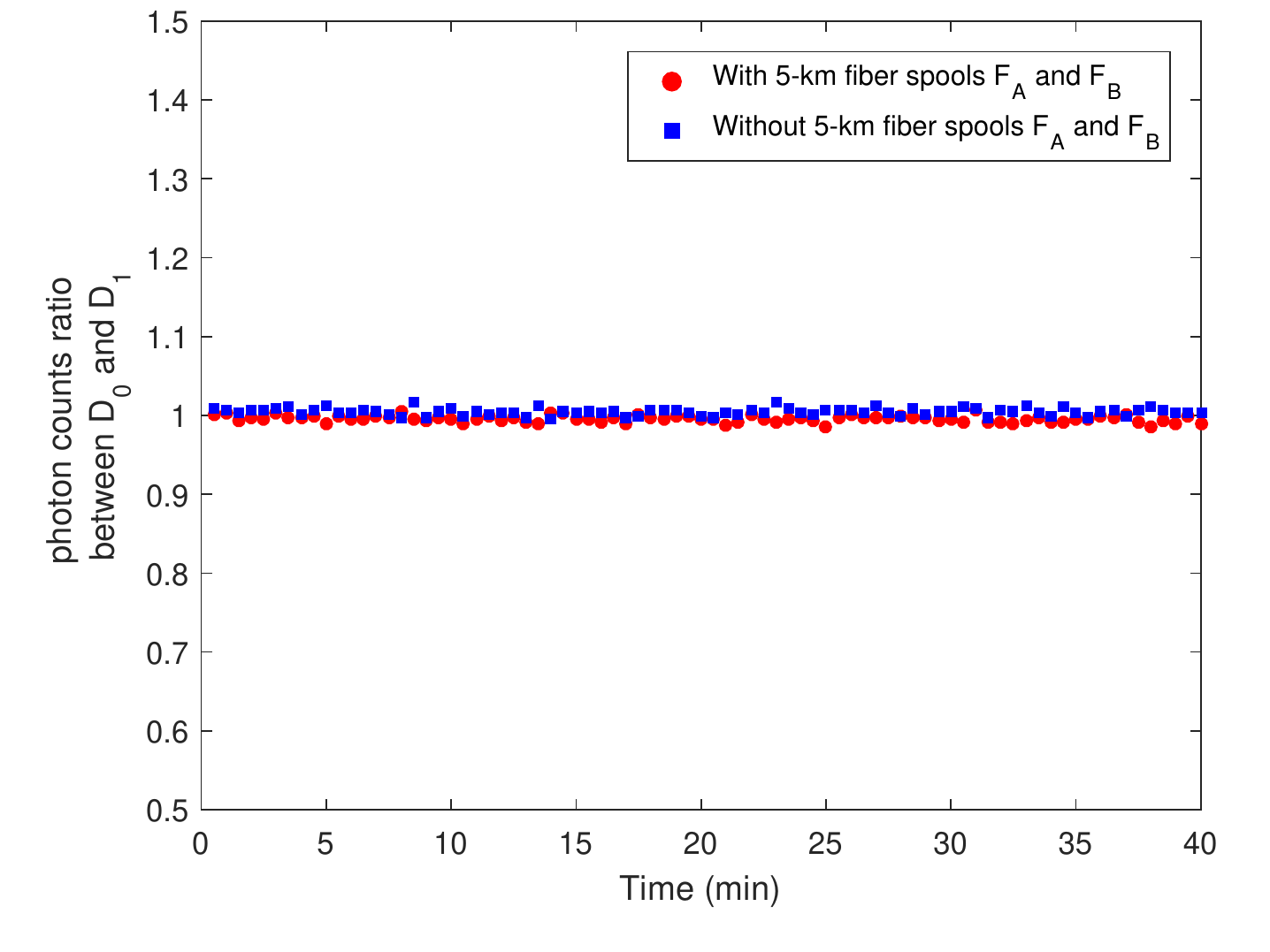}}
	\label{ratio}
	\caption{(a) Interference visibility of the system running in 40 minutes. Each data point represents the average interference visibility in 30s. The blue squares indicate the case in which Alice (Bob) and Charlie are connected only through an attenuator. While the red circles indicate the case in which a 5-km fiber spool F$_A$ (F$_B$) is added between Alice (Bob) and Charlie. For both cases, the system is stable and the interference visibility is kept above 99.5\%. (b) Ratio of the total number of photon counts at detector $D_0$ to the total number of photon counts at detector $D_1$. Phase randomization is applied in 40 minutes. The total number of photon counts at detector $D_0$ is calibrated to compensated the loss in the circular. In this scenario, the photons should be detected by the detectors $D_0$ and $D_1$ with equal probability. Again, the blue squares (red circles) represent the ratio without (with) 5-km fiber spool F$_A$ (F$_B$) inserted between Alice (Bob) and Charlie. As expected, the ratio in both cases is very close to 1.} 
	\label{stablity}	
\end{figure*}

A number of fiber-based polarization controllers are installed inside the Sagnac loop (see Fig.~\ref{set-up}) to ensure that the pulses are aligned in polarization after they travel through the entire loop and interfere at Charlie’s beam splitter. All the fiber spools are stored in sealed boxes, but no active polarization stabilization is applied. The sealed boxes mimic the environment of buried fiber cables. The voltage of the driving signal of Alice’s (Bob’s) PM$_A$ (PM$_B$) is tuned to V$_\pi$ (4.5V) to maximize the interference visibility. Since the two pulses coming from Alice and Bob travel exactly the same path, they undergo the same phase drift and their relative phase is therefore stable without active phase stabilization. Due to the combination of the aforementioned measures, even after traveling kilometers of fiber, the interference visibility of our system is kept well above 99\% for an extensive duration, as shown in Fig.~\ref{stablity} (a). When the fiber spool F$_A$ (F$_B$) is taken out of the loop and Alice (Bob) and Charlie is connected only through an attenuator (blue squares in Fig.~\ref{stablity} (a)), the system is more stable and the average interference visibility is about 99.8\%. When the 5-km fiber spool F$_A$ (F$_B$) is added between Alice (Bob) and Charlie (red circles in Fig.~\ref{stablity} (a)), the interference visibility is slightly lower compared to the cases without the fibers spools F$_A$ and F$_B$. We attribute this degradation of the visibility to polarization fluctuations, the depolarization effect, as well as low-levels of Rayleigh backscattering in long fiber spools. Nonetheless, the interference visibility in this latter case is still stable for at least 40 minutes and the average value is about 99.7\%. In order to keep the high visibility and improve the performance of the system, we stop the experiment every 40 minutes for polarization realignment. When a random phase is required (for the decoy state signal), the voltage of the driving signal is randomly chosen from $0$ to V$_\pi$. In this scenario, the photons should be detected by the detectors $D_0$ and $D_1$ with equal probability. Fig.~\ref{stablity} (b) shows the ratio of the total number of photon counts at $D_0$ to the total number of photon counts at $D_1$ when phase randomization is applied continuously at both Alice’s and Bob’s stations for 40 minutes. Note that the total number of photon counts at detector $D_0$ is calibrated to compensate for the circulator loss. As illustrated in Fig.~\ref{stablity} (b), both cases (with or without the 5-km fiber spools F$_A$ and F$_B$) maintain a stable ratio close to 1, which indicates that phase randomization can be effectively implemented.

\textit{Results and Discussion} - We implement the experiment for four different values of the overall system loss between Alice and Bob, 38.0 dB, 46.7 dB, 55.1 dB and 49.4 dB, respectively. Optical attenuators are applied to simulate the channel loss for the first three system losses. For the 49.4 dB loss, 5-km fiber spools are inserted between Alice (as well as Bob) and Charlie in addition to the attenuator. As discussed above, the detector efficiency is about 11.7\%, which is equivalent to a 9.3 dB loss that we include in the overall system loss. For different values of the system loss, we choose different intensity sets $\left\lbrace |\alpha|^2,\mu,\nu,\omega\right\rbrace $. The selection of the signal intensity $|\alpha|^2$ is done by optimizing a priori the secret key rate formula for a channel model that approximately simulates the expected behavior of an experimental realization, based on the devices’ parameters. In the asymptotic regime, given that the weakest decoy intensity $\omega$ is sufficiently small, the selection of the other two decoy intensities $\mu$ and $\nu$ (within certain limits) is not so crucial and its effect on the resulting secret key rate turns out to be small. So, we take the values of $\mu$ and $\nu$ shown in Table~\ref{tab:intensity} as an example and for experimental convenience. 

\begin{table*}[t]
	\begin{tabular}{cccccc}
		\hline
		\hline
		\multirow{2}{*}{\bf Loss} & \multirow{2}{1.2cm}{\bf Fiber Inserted$^*$ } & \multicolumn{4}{c}{\bf Intensities} 
		\\
		\cline{3-6}& & $|\alpha|^2$ & $\mu$ & $\nu$  & $\omega$  \\ 
		\hline
		$38.0$ dB & No & $0.0256\pm0.0001$ & $0.087\pm0.001$ & $0.0088\pm0.0002$ & $(1.0\pm0.2)\times10^{-4}$ \\
	
		$46.7$ dB & No & $0.02495\pm0.00005$ & $0.0978\pm0.0008$ & $0.0099\pm0.0001$ & $(7.5\pm0.2)\times10^{-5}$ \\
		
		$49.4$ dB & Yes & $0.0183\pm0.0001$ & $0.02005\pm0.00002$ & $0.00828\pm0.00007$ & $(9.2\pm1.0)\times10^{-6}$ \\
	
		$55.1$ dB & No & $0.0175\pm0.0002$ & $0.0382\pm0.0004$ & $0.00790\pm0.00007$ & $(6.5\pm1.0)\times10^{-5}$ \\
		\hline
		\hline
	\end{tabular}
	\caption{\label{tab:intensity} List of intensity sets for the four different values of the overall system loss 38.0 dB, 46.7 dB, 49.4 dB and 55.1 dB. $|\alpha|^2$ is the average photon number (per pulse) of the coherent states in the $X$ basis. $\mu$, $\nu$ and $\omega$ are the average photon number (per pulse) of the decoy states in the $Z$ basis. The uncertainty of each intensity refers to the measurement of its statistical fluctuation. *: 5-km fiber spool F$_A$ (F$_B$) is inserted between Alice (Bob) and Charlie. Note that the fiber spool F$_{A-B}$ is fixed inside the loop for all the four cases to ensure that Alice and Bob are physically separated with kilometers of fiber.}
\end{table*}

\begin{table*}[t]
		\begin{tabular}{ccC{1.3cm}C{1.3cm}C{2.5cm}C{2.5cm}C{2.5cm}C{2.5cm}}
		\hline
		\hline
		\multirow{2}{*}{\bf Loss} & \multirow{2}{*}{\bf Fiber Inserted$^*$ } & \multicolumn{2}{c}{\bf QBER} &  \multicolumn{3}{c}{\bf Experimental Secret Key Rates} &  \multirow{2}{*}{\bf PLOB Bound} 
		\\
		\cline{3-4}\cline{5-7}& & $D_0$ & $D_1$ & $R_{mean}$& $R_{min}$ & $R_{max}$ & \\
		\hline
		$38.0$ dB & No & $0.0032$ & $0.0036$ & $2.6484\times10^{-4}$ & $1.9917\times10^{-4}$ & $3.4765\times10^{-4}$ & $2.2867\times10^{-4}$ \\
	
		$46.7$ dB & No & $0.0058$ & $0.0032$ & $7.8389\times10^{-5}$ & $6.9058\times10^{-5}$ & $8.8458\times10^{-5}$ & $3.0845\times10^{-5}$ \\

		$49.4$ dB & Yes & $0.0059$ & $0.0056$ &  $3.6306\times10^{-5}$ & $2.4061\times10^{-5}$ & $5.4130\times10^{-5}$ & $1.6564\times10^{-5}$ \\
	
		$55.1$ dB & No & $0.0116$ & $0.0108$ &  $1.7542\times10^{-5}$ & $1.0516\times10^{-5}$ & $2.5652\times10^{-5}$ & $4.4584\times10^{-6}$\\
		\hline
		\hline
	\end{tabular}
	\caption{\label{tab:rate} List of experimental results for the four different values of the overall system loss considered, 38.0 dB, 46.7 dB, 49.4 dB and 55.1 dB. QBER is the experimental quantum bit error rate observed at detectors D0 and D1 when Alice and Bob choose the $X$ basis. The experimental secret key rate includes three cases, i.e. the case where intensity fluctuations are disregarded and the worst and best case scenarios where intensity fluctuations are taken into account. These three cases are indicated with the notation R$_{mean}$, R$_{min}$ and R$_{max}$ respectively. Also, for comparison purposes,  this table includes the PLOB bound~\cite{plob} corresponding to each system loss. *: 5-km fiber spool F$_A$ (F$_B$) is inserted between Alice (Bob) and Charlie. Note that the fiber spool F$_{A-B}$ is fixed inside the loop for all the four cases to ensure that Alice and Bob are physically separated with kilometers of fiber.}
\end{table*}

\begin{figure}[t]
	{\includegraphics[width=12.9cm]{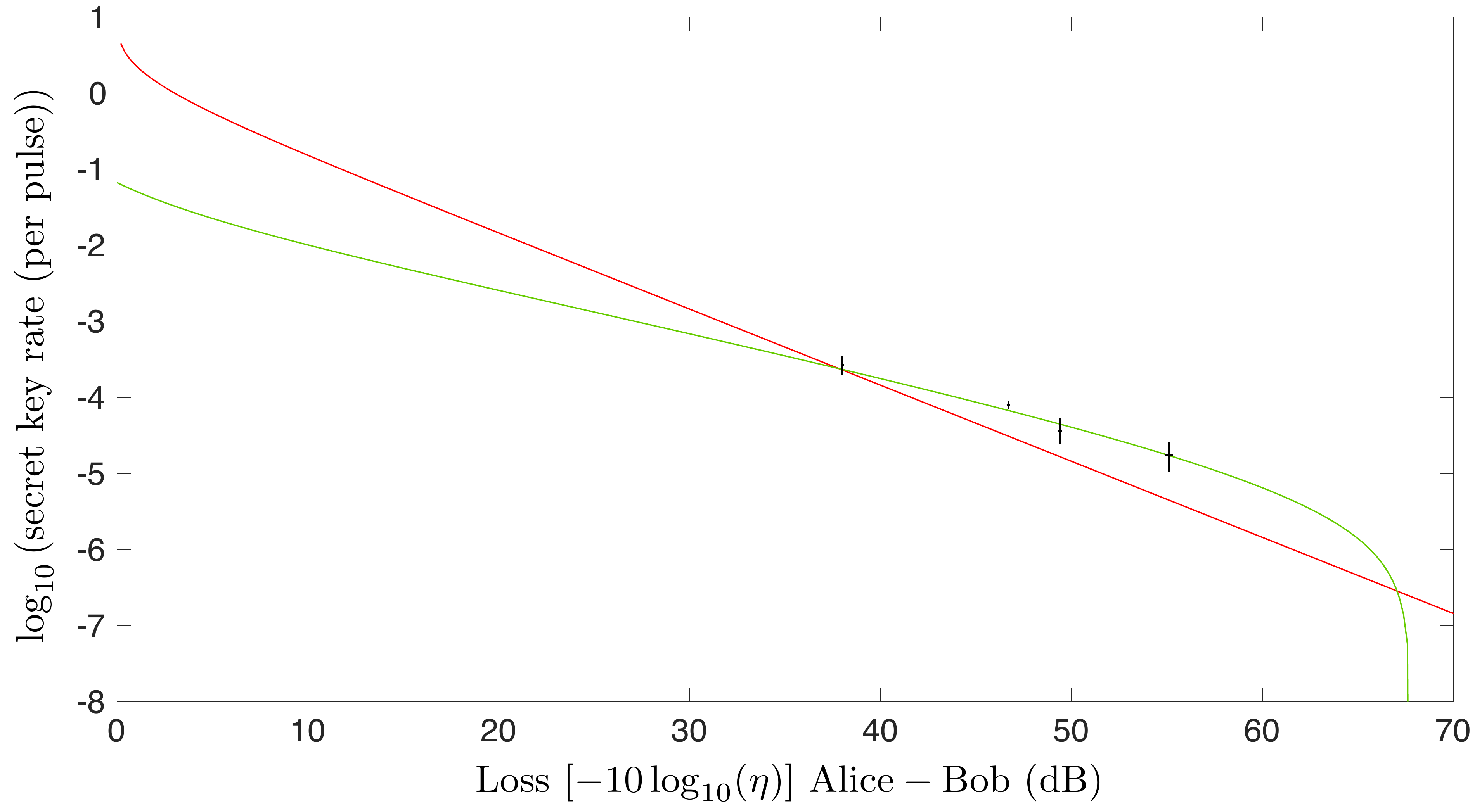}}
	\caption{Secret key rate (per pulse) in logarithmic scale as a function of the overall loss between Alice and Bob. The results shown in Table~\ref{tab:rate} are illustrated as black crosses. The vertical line of each cross shows the difference between the worst and best case scenarios when intensity fluctuations are taken into account, {\it i.e.} the difference between R$_{min}$ and R$_{max}$. The horizontal line of each cross shows the uncertainty of the overall loss, which is $\pm 0.1$ dB for the cases 38.0 dB, 46.7 dB and 49.4 dB system loss, and $\pm 0.2$ for the case 55.1 dB system loss. The point where these two lines cross coincides with the secret key rate if intensity fluctuations are disregarded, {\it i.e.} with R$_{mean}$ in Table~\ref{tab:rate}. The solid red line illustrates the PLOB bound introduced in~\cite{plob}. The solid green line corresponds to a theoretical simulation result realised with the channel model introduced in~\cite{tf-qkd_marcos}. This channel model, for simplicity, assumes the same experimental parameters over all the distances and does not optimise the values of the different decoy intensities which are fixed a priori. Still, the experimental parameters selected are similar to those of the experimental implementation (though these change for each point) and thus also the resulting secret key rate is reasonable similar. Most importantly, our results demonstrate clearly that the experiments performed beat the PLOB
		bound.
		\label{Fig_rates}}
\end{figure}

For each value of the system loss and each intensity pair, Alice and Bob each send out $3\times10^9$ coherent pulses. The experimental quantum bit error rates (QBER) observed in both detectors $D_0$ and $D_1$ when Alice and Bob choose $X$ basis are listed in Table~\ref{tab:rate}. Given the high stability of the system and the high interference visibility, the QBERs observed in the experiment are correspondingly low. Even the maximum QBER observed at the highest system loss is lower than 1.2\%. As the overall system loss decreases, the impact of the dark counts of the single-photon detectors diminishes. This is reflected in the lower QBERs obtained for lower system loss. More experimental data, such as the experimentally observed gains, could be found in the supplementary material~\cite{supplementary}. To extract a secure secret key, we use the security analysis and secret key rate formula reported in~\cite{tf-qkd_marcos} (see supplementary material~\cite{supplementary}). The secret key rates for different system losses, as well as the corresponding PLOB bound, are also listed in Table~\ref{tab:rate}. For each value of the system loss, Table~\ref{tab:rate} includes three cases, i.e. the case where intensity fluctuations are disregarded and the worst- and best-case scenarios where intensity fluctuations are taken into account. For the worst- and best-case scenarios, we numerically minimize and maximize the secret key rate formula among all possible values for the different intensities (within the reported experimental intervals which include the intensity fluctuations). These three cases are indicated in the table with the notation R$_{mean}$, R$_{min}$ and R$_{max}$ respectively. These results are also illustrated in Fig.~\ref{Fig_rates}, which shows the secure key rate (bits per pulse) in logarithmic scale as a function of the overall system loss. The solid red line illustrates the PLOB bound introduced in~\cite{plob}. The solid green line corresponds to a theoretical simulation result realized with the channel model introduced in ~\cite{tf-qkd_marcos}. This channel model, for simplicity, assumes the same experimental parameters (similar to those of the experimental implementation) over all the distances and does not optimize the values of the different decoy intensities which are fixed a priori. The experimental secret key rates are illustrated as black crosses. The vertical line of each cross shows the difference between the worst- and best-case scenarios when intensity fluctuations are taken into account, i.e. the difference between and R$_{min}$ and R$_{max}$. The horizontal line of each cross shows the uncertainty of the overall loss. The point where these two lines cross coincides with the secret key rate if intensity fluctuations are disregarded, i.e. with R$_{mean}$ in Table~\ref{tab:rate}. As depicted in Fig.~\ref{Fig_rates}, the experimental secret key rates are reasonably close to the theoretical simulation results, except that the key rate at the system loss 49.4 dB is slightly lower compared with the simulation result. This is because of the 5-km fiber spools F$_A$ (F$_B$) that are added in between Alice (Bob) and Charlie in this case, which reduces the performance of the system. Nonetheless, the experimental results, as expected, follow the rate-loss dependence of TF-QKD, scaling with the square-root of the channel transmittance. When the overall system loss between Alice and Bob is around 38 dB, the secret key rate sits around the PLOB bound. However, as the overall system loss increases, the experimental key rate evidently surpasses the PLOB bound even when the minimum key rate in the worst-case scenario is considered. This achievement experimentally proves that TF-QKD can beat the fundamental bounds on the private capacity of point-to-point QKD. Particularly, the observed higher key rate, compared with the PLOB bound, at loss 49.4 dB with real fiber spools (F$_A$ and F$_B$) shows the feasibility of the practical application of TF-QKD.

\textit{Conclusion} - In summary, we have implemented the first twin-field quantum key distribution experiment over 10 km of real optical fibers without active phase stabilization or phase post-selection. The secure key rate of the system scales as the square-root of the overall channel transmittance. In particular, we have observed that the resulting secret key rate at the high loss region clearly beats the PLOB bound even when intensity fluctuations are taken into account. Our work shows the feasibility of overcoming the private capacity of a point-to-point QKD link with current technology. Longer fibers could be added into our system in the future to extend the range of TF-QKD.

\textit{Acknowledgments} - We thank Olinka Bedroya and Shihan Sajeed for their assistance and enlightening discussions. We also thank financial support from NSERC, CFI, ORF, US Office of Naval Research, MITACS, Royal Bank of Canada and Huawei Technologies Canada, Ltd. M.C. acknowledges support from the Spanish Ministry of Economy and Competitiveness (MINECO), the Fondo Europeo de Desarrollo Regional (FEDER) through grants TEC2014-54898-R and TEC2017-88243-R, and the European Union’s Horizon 2020 research and innovation programme under the Marie Sklodowska-Curie grant agreement No 675662 (project QCALL). 

%

%%%%%%%%%%%%%%%%%%%%%%%%%%%%%%%%%%%%%%%%%%%%%%%%%%%%%%%%%%%%%%%%%%%%
%\bibliographystyle{ieeetr}
%\bibliographystyle{unsrt}
% \bibliographystyle{apsrev}
%\bibliographystyle{apsrev4-1}
%\bibliography{FiniteHD}

\end{document}

% title
\title{Supplementary Information: Proof-of-principle experimental demonstration of twin-field type quantum key distribution}
\author{Xiaoqing Zhong}
\email{xzhong@physics.utoronto.ca}
\affiliation{Center for Quantum Information and Quantum Control, Dept. of Physics, University of Toronto, Toronto, Ontario, M5S 1A7, Canada}
\author{Jianyong Hu}
\email{jianyong\_hu@163.com}
\affiliation{Center for Quantum Information and Quantum Control, Dept. of Electrical \& Computer Engineering, University of Toronto, Toronto, Ontario, M5S 3G4, Canada}
\affiliation{State Key Laboratory of Quantum Optics and Quantum Optics Devices, Institute of Laser Spectroscopy, Shanxi University, Taiyuan, 030006, China}
\author{Marcos Curty}
\affiliation{EI Telecomunicaci\'on, Dept. of Signal Theory and Communications, University of Vigo, E-36310 Vigo, Spain}
\author{Li Qian}
\affiliation{Center for Quantum Information and Quantum Control, Dept. of Electrical \& Computer Engineering, University of Toronto, Toronto, Ontario, M5S 3G4, Canada}
\author{Hoi-Kwong Lo}
\affiliation{Center for Quantum Information and Quantum Control, Dept. of Physics, University of Toronto, Toronto, Ontario, M5S 1A7, Canada}
\affiliation{Center for Quantum Information and Quantum Control, Dept. of Electrical \& Computer Engineering, University of Toronto, Toronto, Ontario, M5S 3G4, Canada}

%\date{\today}
%%%%%%%%%%%%%%%%%%%%%%%%%%%%%%%%%%%%%%%%%%%%%%%%%%%%%%%%%%%%%%%%%%%%%%%
% Abstract
%%%%%%%%%%%%%%%%%%%%%%%%%%%%%%%%%%%%%%%%%%%%%%%%%%%%%%%%%%%%%%%%%%%%%%%
%\begin{abstract}
%\end{abstract}
\maketitle

%%%%%%%%%%%%%%%%%%%%%%%%%%%%%%%%%%%%%%%%%%%%%%%%%%%%%%%%%%%%%%%%%%%%%%%%%%%%%%%%%%%%%%%

\section{Secret key rate formula}

We use the security analysis introduced in~\cite{tf-qkd_protocol}. It states that a lower bound on the asymptotic secret key rate, $R$, can be written as
\begin{equation}\label{key_rate}
R\geq R_{10}+R_{01},
\end{equation}
where the terms $R_{D_0D_1}$, with $(D_0,D_1)\in\{(1,0),(0,1)\}$, have the form
\begin{eqnarray}\label{qwe5}
R_{D_0D_1}&=&\max{\left\{p_X^2p(D_0,D_1) \left[1-f_{\rm EC}*h(e_{D_0D_1})-h(e^{\rm ph}_{D_0D_1})\right],0\right\}}.
\end{eqnarray}
The term $p_X$ denotes the probability that Alice (Bob) selects the $X$ basis. In the asymptotic regime we assume that this probability is very close to one and thus $p_X^2\approx 1$~\cite{tf-qkd_protocol}. The parameter $p(D_0,D_1)$, on the other hand, represents the conditional probability that Charlie announces the measurement outcome $(D_0,D_1)$ given that both Alice and Bob emit a signal state encoded in the $X$ basis. The case $(D_0,D_1)=(1,0)$ corresponds to observing a ``click'' only in detector $D_0$, while the case $(D_0,D_1)=(0,1)$ corresponds to observing a ``click'' only in detector $D_1$ at Charlie. This probability can be expressed as
\begin{equation}\label{qwe1}
p(D_0,D_1)=\sum_{b_A,b_B=0,1} p(b_A,b_B)p(D_0,D_1|b_A,b_B)=\frac{1}{4}\sum_{b_A,b_B=0,1} p(D_0,D_1|b_A,b_B),
\end{equation}
where $b_A=0$ ($b_A=1$) refers to Alice emitting the coherent state $\ket{\alpha}$ ($\ket{-\alpha}$), and similarly for $b_B$. That is, $p(b_A,b_B)$ denotes the joint probability that Alice emits the signal state associated to $b_A$ and Bob that associated to $b_B$. In Eq.~(\ref{qwe1}) we have used the fact that, in the experiment, $p(b_A,b_B)=1/4$ $\forall b_A, b_B$. Similarly, $p(D_0,D_1|b_A,b_B)$ denotes the conditional probability that Charlie announces the measurement outcome $(D_0,D_1)$ given that Alice and Bob emit, respectively, a signal state associated to $b_A$ and $b_B$. These probabilities can be estimated directly from the experimental data.

In Eq.~(\ref{qwe5}), $f_{EC}$ is an inefficiency function for the error correction process. We fix $f_{EC}$ to a typical value, say $f_{EC}=1.16$~\cite{tf-qkd_protocol}. The function $h(x) = -x \log_2(x) - (1 - x) \log_2(1 - x)$ represents the binary Shannon entropy function, and the term $e_{D_0D_1}$ denotes the quantum bit-error rate in the $X$ basis. This last quantity can be expressed as
\begin{eqnarray}\label{error}
e_{10}&=&\frac{\sum_{i,i|i\oplus j=1}p(D_0=1,D_1=0|b_A=i,b_B=j)}{4p(D_0=1,D_1=0)}, \nonumber\\
e_{01}&=&\frac{\sum_{i=0,1}p(D_0=0,D_1=1|b_A=i,b_B=i)}{4p(D_0=0,D_1=1)},
\end{eqnarray}
where, again, we have used the fact that $p(b_A,b_B)=1/4$ $\forall b_A, b_B$.

Finally, the quantity $e^{\rm ph}_{D_0D_1}$ which appears in Eq.~(\ref{qwe5}) refers to an upper bound on the phase-error rate associated to the signals in the $X$ basis. This quantity satisfies~\cite{tf-qkd_protocol}
\begin{eqnarray}\label{phase}
e^{\rm ph}_{D_0D_1}&\leq&\frac{1}{p(D_0,D_1)}\Bigg\{\Bigg[c_{0}^2\sqrt{Y_{00,D_0D_1}^{\rm U}}+c_{0}c_{2}\Big(\sqrt{Y_{02,D_0D_1}^{\rm U}}+\sqrt{Y_{20,D_0D_1}^{\rm U}}\Big)+\Delta\Bigg]^2
+\Bigg[c_{1}^2\sqrt{Y_{11,D_0D_1}^{\rm U}}+{\bar \Delta}\Bigg]^2\Bigg\},
\end{eqnarray}
where the coefficients $c_i=e^{-\frac{|\alpha|^2}{2}}\alpha^i/\sqrt{i!}$, with $\alpha$ being the amplitude of the $X$ basis signals, and where the residual parameters $\Delta$ and ${\bar \Delta}$ are given, respectively, by
\begin{eqnarray}
\Delta&=&\sum_{n,m=0}^\infty c_{2n}c_{2m}-c_{0}^2-2c_{0}c_{2}, \nonumber \\
{\bar \Delta}&=&\sum_{n,m=0}^\infty c_{2n+1}c_{2m+1}-c_{1}^2.
\end{eqnarray}

The quantities $Y_{nm,D_0D_1}^{\rm U}$, with $(n,m)\in\{(0,0), (0,2), (2,0), (1,1)\}$, that appear in Eq.~(\ref{phase}) refer to an upper bound on the yields $Y_{nm,D_0D_1}$, which are the conditional probabilities that Charlie announces the measurement outcome $(D_0,D_1)$ given that Alice and Bob emit an $n$-photon state and an $m$-photon state, respectively. We estimate these quantities by using the experimental data from the Z basis. 

\subsection{Estimation of $Y_{nm,D_0D_1}^{\rm U}$}

The parameters $Y_{nm,D_0D_1}^{\rm U}$ can be estimated using either analytical or numerical tools. In this section we introduce an analytical method that we use to estimate these quantities. Our starting point is the experimentally observed gains, $Q^{ab}_{D_0,D_1}$, of the $Z$ basis states, which are the conditional probabilities that Charlie announces the measurement outcome $(D_0,D_1)$ given that Alice (Bob) sent him a phase-randomised coherent state of intensity $a$ ($b$), with $a,b\in\mathcal{S}=\{\mu,\nu,\omega\}$. We have that
\begin{equation}
Q^{ab}_{D_0,D_1}=e^{-a-b}\sum_{n,m=0}^\infty \frac{a^nb^m}{n!m!}Y_{nm,D_0D_1}.
\end{equation}

For convenience and in order to simplify the notation, below we shall denote the quantities $Q^{ab}_{D_0,D_1}$ and $Y_{nm,D_0D_1}$ as $Q^{ab}$ and $Y_{nm}$, respectively. That is, we will omit the dependence of these parameters with the detection pattern $(D_0,D_1)$. Likewise, we shall use $Y_{nm}^{{\rm U}}$ instead of $Y_{nm,D_0D_1}^{\rm U}$.

\subsubsection{Estimation of $Y_{00}^{{\rm U}}$}

To upper bound the yield $Y_{00}$ we use the result introduced in~\cite{finite_decoy_tfqkd}. It states that 
\begin{equation}\label{y00}
Y_{00} \leq Y_{00}^{\rm U}\equiv\min\left\{\frac{\frac{\omega^2\Delta^{\mu\nu}}{\mu-\nu}-\frac{\nu^2\Delta^{\mu\omega}}{\mu-\omega}+\frac{\mu^2\Delta^{\nu\omega}}{\nu-\omega}}{(\mu-\nu)(\mu-\omega)(\nu-\omega)}, 1\right\}, \quad \quad
\end{equation}
where the terms $\Delta^{\mu\nu}$, $\Delta^{\mu\omega}$ and $\Delta^{\nu\omega}$ have the form
\begin{eqnarray}
\Delta^{\mu\nu}&=\nu^2e^{2\mu}Q^{\mu\mu}+\mu^2e^{2\nu}Q^{\nu\nu}-\mu\nu e^{\mu+\nu}(Q^{\mu\nu}+Q^{\nu\mu}), \nonumber \\
\Delta^{\mu\omega}&=\omega^2e^{2\mu}Q^{\mu\mu}+\mu^2e^{2\omega}Q^{\omega\omega}-\mu\omega e^{\mu+\omega}(Q^{\mu\omega}+Q^{\omega\mu}), \nonumber \\
\Delta^{\nu\omega}&=\omega^2e^{2\nu}Q^{\nu\nu}+\nu^2e^{2\omega}Q^{\omega\omega}-\nu\omega e^{\nu+\omega}(Q^{\nu\omega}+Q^{\omega\nu}).
\end{eqnarray}

\subsubsection{Estimation of $Y_{11}^{\rm U}$}

To estimate $Y_{11}^{\rm U}$ we proceed as follows. Let $a_1,a_0\in{\mathcal S}$, with $a_1>a_0$, denote Alice's intensities, and let $b_1,b_0\in{\mathcal S}$, with $b_1>b_0$, denote Bob's intensities. Then, 
we have that 
\begin{eqnarray}\label{main_Y11}
\Gamma^{a_1a_0b_1b_0}&&\equiv{}e^{a_0+b_0}Q^{a_0b_0}+e^{a_1+b_1}Q^{a_1b_1}-e^{a_0+b_1}Q^{a_0b_1}-e^{a_1+b_0}Q^{a_1b_0}=\sum_{n,m=0}^\infty\frac{(a_1^n-a_0^n)(b_1^m-b_0^m)}{n!m!}Y_{nm}.
\end{eqnarray}
Since $a_1>a_0$ and $b_1>b_0$, we have that all the terms that multiply the yields $Y_{nm}$ in Eq.~(\ref{main_Y11}) are positive. This means, in particular, that
\begin{equation}
\Gamma^{a_1a_0b_1b_0}\geq (a_1-a_0)(b_1-b_0)Y_{11},
\end{equation}
which implies
\begin{equation}
Y_{11}\leq \min\left\{\frac{\Gamma^{a_1a_0b_1b_0}}{(a_1-a_0)(b_1-b_0)},1\right\}\equiv Y_{U,11}.
\end{equation}

In addition, let ${\bar a}_1,{\bar b}_1\in{\mathcal S}$, with $a_1>{\bar a}_1>a_0$ and $b_1>{\bar b}_1>b_0$. Then,  we have that
\begin{eqnarray}\label{wer}
\Xi^{a_1a_0b_1b_0{\bar a}_1{\bar b}_1}&&\equiv{}({\bar a}_1^2-a_0^2)({\bar b}_1^2-b_0^2)\left[e^{a_1+b_1}Q^{a_1b_1}-e^{a_1+b_0}Q^{a_1b_0}-e^{a_0+b_1}Q^{a_0b_1}+e^{a_0+b_0}Q^{a_0b_0}\right]\nonumber \\
&&-({\bar a}_1^2-a_0^2)(b_1^2-b_0^2)\left[e^{a_1+{\bar b}_1}Q^{a_1{\bar b}_1}-e^{a_1+b_0}Q^{a_1b_0}-e^{a_0+{\bar b}_1}Q^{a_0{\bar b}_1}+e^{a_0+b_0}Q^{a_0b_0}\right]\nonumber \\
&&-(a_1^2-a_0^2)({\bar b}_1^2-b_0^2)\left[e^{{\bar a}_1+b_1}Q^{{\bar a}_1b_1}-e^{{\bar a}_1+b_0}Q^{{\bar a}_1b_0}-e^{a_0+b_1}Q^{a_0b_1}+e^{a_0+b_0}Q^{a_0b_0}\right]\nonumber \\
&&+(a_1^2-a_0^2)(b_1^2-b_0^2)\left[e^{{\bar a}_1+{\bar b}_1}Q^{{\bar a}_1{\bar b}_1}-e^{{\bar a}_1+b_0}Q^{{\bar a}_1b_0}-e^{a_0+{\bar b}_1}Q^{a_0{\bar b}_1}+e^{a_0+b_0}Q^{a_0b_0}\right]\nonumber \\
&&=\sum_{n,m=0}^\infty\frac{[({\bar a}_1^2-a_0^2)(a_1^n-a_0^n)-(a_1^2-a_0^2)({\bar a}_1^n-a_0^n)][({\bar b}_1^2-b_0^2)(b_1^m-b_0^m)-(b_1^2-b_0^2)({\bar b}_1^m-b_0^m)]}{n!m!}Y_{nm}.\quad \quad
\end{eqnarray}
From Eq.~(\ref{wer}) we have that $\Xi^{a_1a_0b_1b_0{\bar a}_1{\bar b}_1}$ can be expressed as
\begin{eqnarray}\label{eqa}
\Xi^{a_1a_0b_1b_0{\bar a}_1{\bar b}_1}&&=[({\bar a}_1^2-a_0^2)(a_1-a_0)-(a_1^2-a_0^2)({\bar a}_1-a_0)][({\bar b}_1^2-b_0^2)(b_1-b_0)-(b_1^2-b_0^2)({\bar b}_1-b_0)]Y_{11}\nonumber \\
&&+[({\bar a}_1^2-a_0^2)(a_1-a_0)-(a_1^2-a_0^2)({\bar a}_1-a_0)]\sum_{m\geq 3}^\infty \frac{[({\bar b}_1^2-b_0^2)(b_1^m-b_0^m)-(b_1^2-b_0^2)({\bar b}_1^m-b_0^m)]}{m!}Y_{1m}\nonumber \\
&&+[({\bar b}_1^2-b_0^2)(b_1-b_0)-(b_1^2-b_0^2)({\bar b}_1-b_0)]\sum_{n\geq 3}^\infty \frac{[({\bar a}_1^2-a_0^2)(a_1^n-a_0^n)-(a_1^2-a_0^2)({\bar a}_1^n-a_0^n)]}{n!}Y_{n1}\nonumber \\
&&+\sum_{n,m\geq 3}^\infty\frac{[({\bar a}_1^2-a_0^2)(a_1^n-a_0^n)-(a_1^2-a_0^2)({\bar a}_1^n-a_0^n)][({\bar b}_1^2-b_0^2)(b_1^m-b_0^m)-(b_1^2-b_0^2)({\bar b}_1^m-b_0^m)]}{n!m!}Y_{nm}.\quad \quad
\end{eqnarray}
Importantly, since $a_1>{\bar a}_1>a_0$, we have that $({\bar a}_1^2-a_0^2)(a_1-a_0)-(a_1^2-a_0^2)({\bar a}_1-a_0)\leq 0$ and $[({\bar a}_1^2-a_0^2)(a_1^n-a_0^n)-(a_1^2-a_0^2)({\bar a}_1^n-a_0^n)]\geq{}0$ for all $n\geq 3$.  

To see this, let us start by considering the first inequality, {\it i.e.} $({\bar a}_1^2-a_0^2)(a_1-a_0)-(a_1^2-a_0^2)({\bar a}_1-a_0)\leq 0$. This inequality is equivalent to
\begin{equation}
\frac{({\bar a}_1^2-a_0^2)}{(a_1^2-a_0^2)}=\frac{({\bar a}_1-a_0)({\bar a}_1+a_0)}{(a_1-a_0)(a_1+a_0)}\leq\frac{({\bar a}_1-a_0)}{(a_1-a_0)}.
\end{equation}
This implies that
\begin{equation}
\frac{({\bar a}_1+a_0)}{(a_1+a_0)}\leq1,
\end{equation}
which is true because $a_1>{\bar a}_1$. 

Let us consider now the second inequality, {\it i.e.} $[({\bar a}_1^2-a_0^2)(a_1^n-a_0^n)-(a_1^2-a_0^2)({\bar a}_1^n-a_0^n)]\geq{}0$ for all $n\geq 3$. This statement is equivalent to
\begin{equation}\label{qwe}
\frac{({\bar a}_1^2-a_0^2)}{(a_1^2-a_0^2)}=\frac{({\bar a}_1-a_0)({\bar a}_1+a_0)}{(a_1-a_0)(a_1+a_0)}\geq \frac{({\bar a}_1^n-a_0^n)}{(a_1^n-a_0^n)}=\frac{({\bar a}_1-a_0)({\bar a}_1^{n-1}+{\bar a}_1^{n-2}a_0+
{\bar a}_1^{n-3}a_0^2+\ldots+{\bar a}_1^2a_0^{n-3}+{\bar a}_1a_0^{n-2}+a_0^{n-1})}{(a_1-a_0)(a_1^{n-1}+a_1^{n-2}a_0+a_1^{n-3}a_0^2+\ldots+a_1^2a_0^{n-3}+a_1a_0^{n-2}+a_0^{n-1})},
\end{equation}
where, on the RHS of the inequality given by Eq.~(\ref{qwe}), we have used the fact that $x^n-y^n=(x-y)(x^{n-1}+x^{n-2}y+x^{n-3}y^2+\ldots+x^2y^{n-3}+xy^{n-2}+y^{n-1})$. This means that Eq.~(\ref{qwe}) can be written as
\begin{eqnarray}\label{ty7}
\frac{({\bar a}_1+a_0)}{(a_1+a_0)}\geq\frac{({\bar a}_1^{n-1}+{\bar a}_1^{n-2}a_0+
{\bar a}_1^{n-3}a_0^2+\ldots+{\bar a}_1^2a_0^{n-3}+{\bar a}_1a_0^{n-2}+a_0^{n-1})}{(a_1^{n-1}+a_1^{n-2}a_0+a_1^{n-3}a_0^2+\ldots+a_1^2a_0^{n-3}+a_1a_0^{n-2}+a_0^{n-1})}.
\end{eqnarray}
After some straightforward calculations we obtain that Eq.~(\ref{ty7}) implies
\begin{eqnarray}\label{qwer}
({\bar a}_1a_1^2+a_0a_1^2)(a_1^{n-3}+a_1^{n-4}a_0+a_1^{n-5}a_0^2+\ldots+a_1a_0^{n-4}+a_0^{n-3}) &\geq& 
(a_1{\bar a}_1^2+a_0{\bar a}_1^2)({\bar a}_1^{n-3}+{\bar a}_1^{n-4}a_0+{\bar a}_1^{n-5}a_0^2 \nonumber \\
&+&\ldots+{\bar a}_1a_0^{n-4}+a_0^{n-3}). \quad \quad
\end{eqnarray}
Then, since $a_1>{\bar a}_1$, we find that $({\bar a}_1a_1^2+a_0a_1^2)\geq (a_1{\bar a}_1^2+a_0{\bar a}_1^2)$ and $(a_1^{n-3}+a_1^{n-4}a_0+a_1^{n-5}a_0^2+\ldots+a_1a_0^{n-4}+a_0^{n-3})\geq ({\bar a}_1^{n-3}+{\bar a}_1^{n-4}a_0+{\bar a}_1^{n-5}a_0^2+\ldots+{\bar a}_1a_0^{n-4}+a_0^{n-3})$, which confirms that Eq.~(\ref{qwer}) also holds. 

From the results above it follows directly that $({\bar b}_1^2-b_0^2)(b_1-b_0)-(b_1^2-b_0^2)({\bar b}_1-b_0)\leq 0$ and $[({\bar b}_1^2-b_0^2)(b_1^m-b_0^m)-(b_1^2-b_0^2)({\bar b}_1^m-b_0^m)]\geq{}0$ for all $m\geq 3$ whenever $b_1>{\bar b}_1>b_0$.

This means that in Eq.~(\ref{eqa}) the terms that multiply the yields $Y_{11}$ and $Y_{nm}$ with $n,m\geq 3$ are greater or equal to zero, while the terms that multiply the yields $Y_{1m}$ and $Y_{n1}$ with $n,m\geq 3$ are smaller or equal to zero. We obtain, therefore, that the parameter $\Xi^{a_1a_0b_1b_0{\bar a}_1{\bar b}_1}$ is lower bounded by
\begin{eqnarray}
\Xi^{a_1a_0b_1b_0{\bar a}_1{\bar b}_1}&&\geq [({\bar a}_1^2-a_0^2)(a_1-a_0)-(a_1^2-a_0^2)({\bar a}_1-a_0)][({\bar b}_1^2-b_0^2)(b_1-b_0)-(b_1^2-b_0^2)({\bar b}_1-b_0)]Y_{11}+\zeta^{a_1a_0b_1b_0{\bar a}_1{\bar b}_1}, \nonumber \\
\end{eqnarray}
where the quantity $\zeta^{a_1a_0b_1b_0{\bar a}_1{\bar b}_1}$ is given by
\begin{eqnarray}
\zeta^{a_1a_0b_1b_0{\bar a}_1{\bar b}_1}&&=[({\bar a}_1^2-a_0^2)(a_1-a_0)-(a_1^2-a_0^2)({\bar a}_1-a_0)]\sum_{m\geq 3}^\infty \frac{[({\bar b}_1^2-b_0^2)(b_1^m-b_0^m)-(b_1^2-b_0^2)({\bar b}_1^m-b_0^m)]}{m!}\nonumber \\
&&+[({\bar b}_1^2-b_0^2)(b_1-b_0)-(b_1^2-b_0^2)({\bar b}_1-b_0)]\sum_{n\geq 3}^\infty \frac{[({\bar a}_1^2-a_0^2)(a_1^n-a_0^n)-(a_1^2-a_0^2)({\bar a}_1^n-a_0^n)]}{n!} \nonumber \\
&&=[({\bar a}_1^2-a_0^2)(a_1-a_0)-(a_1^2-a_0^2)({\bar a}_1-a_0)]\bigg\{({\bar b}_1^2-b_0^2)\bigg[b_0+\frac{b_0^2}{2}-\frac{b_1}{2}(2+b_1)-e^{b_0}+e^{b_1}\bigg] \nonumber \\
&&-(b_1^2-b_0^2)\bigg[b_0+\frac{b_0^2}{2}-\frac{{\bar b}_1}{2}(2+{\bar b}_1)-e^{b_0}+e^{{\bar b}_1}\bigg]\bigg\}+[({\bar b}_1^2-b_0^2)(b_1-b_0)-(b_1^2-b_0^2)({\bar b}_1-b_0)] \nonumber \\
&&\bigg\{({\bar a}_1^2-a_0^2)\bigg[a_0+\frac{a_0^2}{2}-\frac{a_1}{2}(2+a_1)-e^{a_0}+e^{a_1}\bigg]
-(a_1^2-a_0^2)\bigg[a_0+\frac{a_0^2}{2}-\frac{{\bar a}_1}{2}(2+{\bar a}_1)-e^{a_0}+e^{{\bar a}_1}\bigg]\bigg\}. \nonumber \\
\end{eqnarray}

We find, therefore, that
\begin{equation}
Y_{11}\leq\min\left\{\frac{\Xi^{a_1a_0b_1b_0{\bar a}_1{\bar b}_1}-\zeta^{a_1a_0b_1b_0{\bar a}_1{\bar b}_1}}{[({\bar a}_1^2-a_0^2)(a_1-a_0)-(a_1^2-a_0^2)({\bar a}_1-a_0)][({\bar b}_1^2-b_0^2)(b_1-b_0)-(b_1^2-b_0^2)({\bar b}_1-b_0)]},1\right\}\equiv Y'_{U,11}.
\end{equation}

This means that 
\begin{equation}\label{y11}
Y_{11}\leq Y_{11}^{\rm U}\equiv\min\left\{Y_{U,11},Y'_{U,11}\right\}.
\end{equation}

\subsubsection{Estimation of $Y_{02}^{\rm U}$}

Let the parameter $\Omega^{a_1a_0b_1b_0{\bar b}_1}$, with $a_1>a_0$ and $b_1>{\bar b}_1>b_0$, be defined as
\begin{eqnarray}\label{eq02}
\Omega^{a_1a_0b_1b_0{\bar b}_1}&&\equiv{}(b_1-{\bar b}_1)\left[a_1e^{a_0+b_0}Q^{a_0b_0}-a_0e^{a_1+b_0}Q^{a_1b_0}\right]+(b_1-b_0)\left[a_0e^{a_1+{\bar b}_1}Q^{a_1{\bar b}_1}-a_1e^{a_0+{\bar b}_1}Q^{a_0{\bar b}_1}\right]\nonumber \\
&&+({\bar b}_1-b_0)\left[a_1e^{a_0+b_1}Q^{a_0b_1}-a_0e^{a_1+b_1}Q^{a_1b_1}\right]\nonumber \\
&&=\sum_{n,m=0}^\infty\frac{(a_0^na_1-a_0a_1^n)\left[({\bar b}_1-b_0)(b_1^m-b_0^m)-(b_1-b_0)({\bar b}_1^m-b_0^m)\right]}{n!m!}Y_{nm} \nonumber \\
&&=\frac{(a_1-a_0)(b_1-b_0)({\bar b}_1-b_0)(b_1-{\bar b}_1)}{2}Y_{02}+(a_1-a_0)\sum_{m\geq 3}^\infty\frac{({\bar b}_1-b_0)(b_1^m-b_0^m)-(b_1-b_0)({\bar b}_1^m-b_0^m)}{m!}Y_{0m} \nonumber \\
&&+\sum_{n,m\geq 2}^\infty\frac{(a_0^na_1-a_0a_1^n)\left[({\bar b}_1-b_0)(b_1^m-b_0^m)-(b_1-b_0)({\bar b}_1^m-b_0^m)\right]}{n!m!}Y_{nm}.
\end{eqnarray}
Since $a_1>a_0$ and $b_1>{\bar b}_1>b_0$, we have that the terms $({\bar b}_1-b_0)(b_1^m-b_0^m)-(b_1-b_0)({\bar b}_1^m-b_0^m)\geq 0$ for all $m\geq 2$. To demonstrate this, note that this statement is equivalent to
\begin{equation}
\frac{({\bar b}_1-b_0)}{(b_1-b_0)}\geq\frac{({\bar b}_1^m-b_0^m)}{(b_1^m-b_0^m)}=\frac{({\bar b}_1-b_0)({\bar b}_1^{n-1}+{\bar b}_1^{n-2}b_0+
{\bar b}_1^{n-3}b_0^2+\ldots+{\bar b}_1^2b_0^{n-3}+{\bar b}_1b_0^{n-2}+b_0^{n-1})}{(b_1-b_0)(b_1^{n-1}+b_1^{n-2}b_0+b_1^{n-3}b_0^2+\ldots+b_1^2b_0^{n-3}+b_1b_0^{n-2}+b_0^{n-1})}.
\end{equation} 
This implies that
\begin{equation}
1\geq\frac{({\bar b}_1^{n-1}+{\bar b}_1^{n-2}b_0+
{\bar b}_1^{n-3}b_0^2+\ldots+{\bar b}_1^2b_0^{n-3}+{\bar b}_1b_0^{n-2}+b_0^{n-1})}{(b_1^{n-1}+b_1^{n-2}b_0+b_1^{n-3}b_0^2+\ldots+b_1^2b_0^{n-3}+b_1b_0^{n-2}+b_0^{n-1})},
\end{equation} 
which is true because $b_1>{\bar b}_1$.

We find, therefore, that in Eq.~(\ref{eq02}) the terms that multiply the yields $Y_{0m}$ with $m\geq 2$ are greater or equal to zero, while those that multiply the yields $Y_{nm}$ with $n,m\geq 2$ are smaller or equal to zero because the terms $a_0^na_1-a_0a_1^n$ are smaller than zero. This means that
\begin{eqnarray}
\Omega^{a_1a_0b_1b_0{\bar b}_1}&&\geq\frac{(a_1-a_0)(b_1-b_0)({\bar b}_1-b_0)(b_1-{\bar b}_1)}{2}Y_{02} \nonumber \\
&&+\sum_{n,m\geq 2}^\infty\frac{(a_0^na_1-a_0a_1^n)\left[({\bar b}_1-b_0)(b_1^m-b_0^m)-(b_1-b_0)({\bar b}_1^m-b_0^m)\right]}{n!m!}\nonumber \\
&&=\frac{(a_1-a_0)(b_1-b_0)({\bar b}_1-b_0)(b_1-{\bar b}_1)}{2}Y_{02}+\lambda^{a_1a_0b_1b_0{\bar b}_1},
\end{eqnarray}
where the parameter $\lambda^{a_1a_0b_1b_0{\bar b}_1}$ is given by
\begin{equation}
\lambda^{a_1a_0b_1b_0{\bar b}_1}=\left[a_0(1-e^{a_1})-a_1(1-e^{a_0})\right][({\bar b}_1-b_0)(b_0-b_1-e^{b_0}+e^{b_1})-(b_1-b_0)(b_0-{\bar b}_1-e^{b_0}+e^{{\bar b}_1})].
\end{equation}

This implies that
\begin{eqnarray}\label{y02}
Y_{02}\leq Y_{02}^{\rm U}\equiv\min\left\{\frac{2\left[\Omega^{a_1a_0b_1b_0{\bar b}_1}-\lambda^{a_1a_0b_1b_0{\bar b}_1}\right]}{(a_1-a_0)(b_1-b_0)({\bar b}_1-b_0)(b_1-{\bar b}_1)},1\right\}. \quad
\end{eqnarray}

\subsubsection{Estimation of $Y_{20}^{\rm U}$}

Similarly, let the quantity $\Lambda^{a_1a_0b_1b_0{\bar a}_1}$ be defined as
\begin{eqnarray}
\Lambda^{a_1a_0b_1b_0{\bar a}_1}&&\equiv{}(a_1-{\bar a}_1)\left[b_1e^{a_0+b_0}Q^{a_0b_0}-b_0e^{a_0+b_1}Q^{a_0b_1}\right]+(a_1-a_0)\left[b_0e^{{\bar a}_1+b_1}Q^{{\bar a}_1b_1}-b_1e^{{\bar a}_1+b_0}Q^{{\bar a}_1b_0}\right]\nonumber \\
&&+({\bar a}_1-a_0)\left[b_1e^{a_1+b_0}Q^{a_1b_0}-b_0e^{a_1+b_1}Q^{a_1b_1}\right]\nonumber \\
&&=\sum_{n,m=0}^\infty\frac{\left[({\bar a}_1-a_0)(a_1^n-a_0^n)-(a_1-a_0)({\bar a}_1^n-a_0^n)\right](b_0^mb_1-b_0b_1^m)}{n!m!}Y_{nm} \nonumber \\
&&=\frac{(a_1-a_0)({\bar a}_1-a_0)(a_1-{\bar a}_1)(b_1-b_0)}{2}Y_{20}+(b_1-b_0)\sum_{n\geq 3}^\infty\frac{({\bar a}_1-a_0)(a_1^n-a_0^n)-(a_1-a_0)({\bar a}_1^n-a_0^n)}{n!}Y_{n0} \nonumber \\
&&+\sum_{n,m\geq 2}^\infty\frac{\left[({\bar a}_1-a_0)(a_1^n-a_0^n)-(a_1-a_0)({\bar a}_1^n-a_0^n)\right](b_0^mb_1-b_0b_1^m)}{n!m!}Y_{nm}.
\end{eqnarray}

By using exactly the same reasoning employed in the previous section, we have that whenever $a_1>{\bar a}_1>a_0$ and $b_1>b_0$, the terms that multiply the yields $Y_{n0}$ with $n\geq 2$ are greater or equal to zero, while those that multiply the yields $Y_{nm}$ with $n,m\geq 2$ are smaller or equal to zero because the terms $b_0^mb_1-b_0b_1^m$ are smaller than zero. This means that
\begin{eqnarray}
\Lambda^{a_1a_0b_1b_0{\bar a}_1}&&\geq\frac{(a_1-a_0)({\bar a}_1-a_0)(a_1-{\bar a}_1)(b_1-b_0)}{2}Y_{20} \nonumber \\
&&+\sum_{n,m\geq 2}^\infty\frac{\left[({\bar a}_1-a_0)(a_1^n-a_0^n)-(a_1-a_0)({\bar a}_1^n-a_0^n)\right](b_0^mb_1-b_0b_1^m)}{n!m!}\nonumber \\
&&=\frac{(a_1-a_0)({\bar a}_1-a_0)(a_1-{\bar a}_1)(b_1-b_0)}{2}Y_{20}+\tau^{a_1a_0b_1b_0{\bar a}_1},
\end{eqnarray}
where the parameter $\tau^{a_1a_0b_1b_0{\bar a}_1}$ is given by
\begin{equation}
\tau^{a_1a_0b_1b_0{\bar a}_1}=[({\bar a}_1-a_0)(a_0-a_1-e^{a_0}+e^{a_1})-(a_1-a_0)(a_0-{\bar a}_1-e^{a_0}+e^{{\bar a}_1})]\left[b_0(1-e^{b_1})-b_1(1-e^{b_0})\right].
\end{equation}

This implies that
\begin{eqnarray}\label{y20}
Y_{20}\leq Y_{20}^{\rm U}\equiv\min\left\{\frac{2\left[\Lambda^{a_1a_0b_1b_0{\bar a}_1}-\tau^{a_1a_0b_1b_0{\bar a}_1}\right]}
{(a_1-a_0)({\bar a}_1-a_0)(a_1-{\bar a}_1)(b_1-b_0)},1\right\}. 
\end{eqnarray}

\section{Evaluation of the experimental data}

Table~\ref{tab:yields} shows the values of the parameters $p(D_0,D_1)$, $e_{D_0D_1}$, $Y_{nm,D_0D_1}^{\rm U}$, $e^{\rm ph}_{D_0D_1}$ and $R_{D_0D_1}$, with $(D_0,D_1)\in\{(1,0),(0,1)\}$ and $(n,m)\in\{(0,0), (0,2), (2,0), (1,1)\}$, for four different values of the overall system loss, 38.0 dB, 46.7 dB, 49.4 dB and 55.1 dB. As described in the main text, in the case of 49.4 dB system loss, a 5-km fiber spool is inserted between Alice (Bob) and Charlie in addition to the attenuator. 

The parameters $p(D_0,D_1)$ and $e_{D_0D_1}$ are directly obtained from the experimental data in Table~\ref{tab:prob} corresponding to the signals in the X basis by using Eqs.~(\ref{qwe1})-(\ref{error}). The quantities $Y_{nm,D_0D_1}^{\rm U}$, with $(n,m)\in\{(0,0), (0,2), (2,0), (1,1)\}$, are calculated by means of Eqs.~(\ref{y00})-(\ref{y11})-(\ref{y02})-(\ref{y20}) together with the experimentally observed gains $Q^{ab}_{D_0,D_1}$ provided in Table~\ref{tab:gains}, with $a,b\in\mathcal{S}=\{\mu,\nu,\omega\}$, corresponding to the signals in the Z basis. Finally, the phase error rate $e^{\rm ph}_{D_0D_1}$ is obtained by using Eq.~(\ref{phase}), while the secret key rate $R_{D_0D_1}$ is evaluated by using Eq.~(\ref{qwe5}).

For each value of the overall system loss, Table~\ref{tab:yields} considers three different cases. The first one refers to the situation where we disregard the intensity fluctuations of the laser source. That is, the signal intensity $|\alpha|^2$ and the decoy intensities $\mu$, $\nu$ and $\omega$ correspond to the mean values reported in Table I in the main text. The second (third) case refers to the situation where we consider intensity fluctuations and take as signal intensity and decoy intensities those, within the intervals reported in Table I in the main text, that minimise (maximise) the resulting key rate. That is, here we take the worst (best) case scenario that is compatible with our experimental data if intensity fluctuations are taken into account. These three cases are indicated in the table with the notation $R_{D_0D_1}$, $\min_{\mu,\nu,\omega}R_{D_0D_1}$ and $\max_{\mu,\nu,\omega}R_{D_0D_1}$, respectively, and they refer to the quantities $R_{mean}$, $R_{min}$ and $R_{max}$ presented in the main text.

The final secret key rate given by Eq.~(\ref{key_rate}), {\it i.e.} $R_{10}+R_{01}$, is shown in Table~\ref{tab:Key} for the four different values of the overall system loss considered. For each value of the system loss, this Table includes the three cases mentioned above, {\it i.e.} the case where intensity fluctuations are disregarded and the worst and best case scenarios when intensity fluctuations are taken into account. Also, for comparison purposes,  Table~\ref{tab:Key} includes the PLOB bound~\cite{plob}, which reads $-\log_2{(1-\eta)}$, being $\eta$ the system transmittance. These results are also illustrated in Fig.3 in the main text. 

\begin{table}
\begin{tabular}{|c|c|c|c|c|c|c|c|c|}
\hline
{\bf Loss: $38.0$ dB} & $p(D_0,D_1)$ & $e_{D_0D_1}$ &  $Y_{00,D_0D_1}^{\rm U}$ &  $Y_{11,D_0D_1}^{\rm U}$ &  $Y_{02,D_0D_1}^{\rm U}$ &  $Y_{20,D_0D_1}^{\rm U}$ & $e^{\rm ph}_{D_0D_1}$ & $R_{D_0D_1}$ \\
\hline
\hline
$D_0=1$, $D_1=0$ & 3.1823e-04 & 3.2080e-03 & 5.8083e-07 & 8.6573e-03 & 1.9756e-02 & 1.3268e-02 & 1.3273e-01 & 1.2695e-04 \\
\hline
$D_0=0$, $D_1=1$ & 3.1465e-04 & 3.5903e-03 & 4.3238e-07 & 2.6890e-03 & 2.0288e-02 & 1.3618e-02 & 1.1743e-01 & 1.3789e-04 \\
\hline
\hline
{\bf Loss: $38.0$ dB} & $p(D_0,D_1)$ & $e_{D_0D_1}$ &  $Y_{00,D_0D_1}^{\rm U}$ &  $Y_{11,D_0D_1}^{\rm U}$ &  $Y_{02,D_0D_1}^{\rm U}$ &  $Y_{20,D_0D_1}^{\rm U}$ & $e^{\rm ph}_{D_0D_1}$ & $\min_{\mu,\nu,\omega}R_{D_0D_1}$ \\
\hline
\hline
$D_0=1$, $D_1=0$ & 3.1823e-04 & 3.2080e-03 & 8.7112e-07 & 8.6774e-03 & 2.6093e-02 & 1.9702e-02 & 1.7313e-01 & 9.5152e-05 \\
\hline
$D_0=0$, $D_1=1$ & 3.1465e-04 & 3.5903e-03 & 7.2132e-07 & 2.5793e-03 & 2.6566e-02 & 2.0110e-02 & 1.5803e-01 & 1.0402e-04 \\
\hline
\hline
{\bf Loss: $38.0$ dB} & $p(D_0,D_1)$ & $e_{D_0D_1}$ &  $Y_{00,D_0D_1}^{\rm U}$ &  $Y_{11,D_0D_1}^{\rm U}$ &  $Y_{02,D_0D_1}^{\rm U}$ &  $Y_{20,D_0D_1}^{\rm U}$ & $e^{\rm ph}_{D_0D_1}$ & $\max_{\mu,\nu,\omega}R_{D_0D_1}$ \\
\hline
\hline
$D_0=1$, $D_1=0$ & 3.1823e-04 & 3.2080e-03 & 2.7357e-07 & 8.6353e-03 & 1.3446e-02 & 6.8554e-03 & 9.1286e-02 & 1.6645e-04 \\
\hline
$D_0=0$, $D_1=1$ & 3.1465e-04 & 3.5903e-03 & 1.2649e-07 & 2.7937e-03 & 1.4037e-02 & 7.1457e-03 & 7.4989e-02 & 1.8120e-04 \\
\hline
\hline
{\bf Loss: $46.7$ dB} & $p(D_0,D_1)$ & $e_{D_0D_1}$ &  $Y_{00,D_0D_1}^{\rm U}$ &  $Y_{11,D_0D_1}^{\rm U}$ &  $Y_{02,D_0D_1}^{\rm U}$ &  $Y_{20,D_0D_1}^{\rm U}$ & $e^{\rm ph}_{D_0D_1}$ & $R_{D_0D_1}$ \\
\hline
\hline
$D_0=1$, $D_1=0$ & 1.1625e-04 & 5.8043e-03 & 6.9776e-07 & 4.6722e-04 & 6.0460e-03 & 5.6272e-03 & 1.4675e-01 & 3.9369e-05 \\
\hline
$D_0=0$, $D_1=1$ & 1.1614e-04 & 3.1767e-03 & 4.0039e-07 & 2.1468e-03 & 6.7111e-03 & 6.3392e-03 & 1.5744e-01 & 3.9020e-05 \\
\hline
\hline
{\bf Loss: $46.7$ dB} & $p(D_0,D_1)$ & $e_{D_0D_1}$ &  $Y_{00,D_0D_1}^{\rm U}$ &  $Y_{11,D_0D_1}^{\rm U}$ &  $Y_{02,D_0D_1}^{\rm U}$ &  $Y_{20,D_0D_1}^{\rm U}$ & $e^{\rm ph}_{D_0D_1}$ & $\min_{\mu,\nu,\omega}R_{D_0D_1}$ \\
\hline
\hline
$D_0=1$, $D_1=0$ & 1.1625e-04 & 5.8043e-03 & 7.1132e-07 & 4.6995e-04 & 7.1038e-03 & 6.6833e-03 & 1.6413e-01 & 3.4434e-05 \\
\hline
$D_0=0$, $D_1=1$ & 1.1614e-04 & 3.1767e-03 & 4.1421e-07 & 2.1339e-03 & 7.7372e-03 & 7.3834e-03 & 1.7365e-01 & 3.4624e-05 \\
\hline
\hline
{\bf Loss: $46.7$ dB} & $p(D_0,D_1)$ & $e_{D_0D_1}$ &  $Y_{00,D_0D_1}^{\rm U}$ &  $Y_{11,D_0D_1}^{\rm U}$ &  $Y_{02,D_0D_1}^{\rm U}$ &  $Y_{20,D_0D_1}^{\rm U}$ & $e^{\rm ph}_{D_0D_1}$ & $\max_{\mu,\nu,\omega}R_{D_0D_1}$ \\
\hline
\hline
$D_0=1$, $D_1=0$ & 1.1625e-04 & 5.8043e-03 & 6.8392e-07 & 4.6456e-04 & 5.0054e-03 & 4.5882e-03 & 1.2930e-01 & 4.4729e-05 \\
\hline
$D_0=0$, $D_1=1$ & 1.1614e-04 & 3.1767e-03 & 3.8627e-07 & 2.1591e-03 & 5.7019e-03 & 5.3122e-03 & 1.4129e-01 & 4.3729e-05 \\
\hline
\hline
{\bf Loss: $49.4$ dB} & $p(D_0,D_1)$ & $e_{D_0D_1}$ &  $Y_{00,D_0D_1}^{\rm U}$ &  $Y_{11,D_0D_1}^{\rm U}$ &  $Y_{02,D_0D_1}^{\rm U}$ &  $Y_{20,D_0D_1}^{\rm U}$ & $e^{\rm ph}_{D_0D_1}$ & $R_{D_0D_1}$ \\
\hline
\hline
$D_0=1$, $D_1=0$ & 6.3155e-05 & 5.9033e-03 & 6.7859e-07 & 1.7455e-04 & 3.3325e-03 & 1.0892e-02 & 1.6311e-01 & 1.8804e-05 \\
\hline
$D_0=0$, $D_1=1$ & 6.2834e-05 & 5.6327e-03 & 6.2168e-07 & 1.7629e-05 & 4.8162e-03 & 1.0738e-02 & 1.7243e-01 & 1.7502e-05 \\
\hline
\hline
{\bf Loss: $49.4$ dB} & $p(D_0,D_1)$ & $e_{D_0D_1}$ &  $Y_{00,D_0D_1}^{\rm U}$ &  $Y_{11,D_0D_1}^{\rm U}$ &  $Y_{02,D_0D_1}^{\rm U}$ &  $Y_{20,D_0D_1}^{\rm U}$ & $e^{\rm ph}_{D_0D_1}$ & $\min_{\mu,\nu,\omega}R_{D_0D_1}$ \\
\hline
\hline
$D_0=1$, $D_1=0$ & 6.3155e-05 & 5.9033e-03 & 6.8242e-07 & 1.6076e-04 & 6.1573e-03 & 1.3737e-02 & 2.1087e-01 & 1.2397e-05 \\
\hline
$D_0=0$, $D_1=1$ & 6.2834e-05 & 5.6327e-03 & 6.2551e-07 & 2.4009e-06 & 7.6507e-03 & 1.3592e-02 & 2.1780e-01 & 1.1664e-05 \\
\hline
\hline
{\bf Loss: $49.4$ dB} & $p(D_0,D_1)$ & $e_{D_0D_1}$ &  $Y_{00,D_0D_1}^{\rm U}$ &  $Y_{11,D_0D_1}^{\rm U}$ &  $Y_{02,D_0D_1}^{\rm U}$ &  $Y_{20,D_0D_1}^{\rm U}$ & $e^{\rm ph}_{D_0D_1}$ & $\max_{\mu,\nu,\omega}R_{D_0D_1}$ \\
\hline
\hline
$D_0=1$, $D_1=0$ & 6.3155e-05 & 5.9033e-03 & 6.7464e-07 & 1.8810e-04 & 5.0816e-04 & 8.0483e-03 & 1.0493e-01 & 2.8736e-05 \\
\hline
$D_0=0$, $D_1=1$ & 6.2834e-05 & 5.6327e-03 & 6.1773e-07 & 3.2598e-05 & 1.9822e-03 & 7.8855e-03 & 1.2290e-01 & 2.5394e-05 \\
\hline
\hline
{\bf Loss: $55.1$ dB} & $p(D_0,D_1)$ & $e_{D_0D_1}$ &  $Y_{00,D_0D_1}^{\rm U}$ &  $Y_{11,D_0D_1}^{\rm U}$ &  $Y_{02,D_0D_1}^{\rm U}$ &  $Y_{20,D_0D_1}^{\rm U}$ & $e^{\rm ph}_{D_0D_1}$ & $R_{D_0D_1}$ \\
\hline
\hline
$D_0=1$, $D_1=0$ & 3.1489e-05 & 1.1575e-02 & 5.9412e-07 & 2.0453e-04 & 8.1004e-04 & 2.5307e-03 & 1.3355e-01 & 1.0305e-05 \\
\hline
$D_0=0$, $D_1=1$ & 3.1287e-05 & 1.0779e-02 & 5.0726e-07 & 1.2045e-03 & 1.3736e-03 & 3.8711e-03 & 1.7501e-01 & 7.2368e-06 \\
\hline
\hline
{\bf Loss: $55.1$ dB} & $p(D_0,D_1)$ & $e_{D_0D_1}$ &  $Y_{00,D_0D_1}^{\rm U}$ &  $Y_{11,D_0D_1}^{\rm U}$ &  $Y_{02,D_0D_1}^{\rm U}$ &  $Y_{20,D_0D_1}^{\rm U}$ & $e^{\rm ph}_{D_0D_1}$ & $\min_{\mu,\nu,\omega}R_{D_0D_1}$ \\
\hline
\hline
$D_0=1$, $D_1=0$ & 3.1489e-05 & 1.1575e-02 & 6.1349e-07 & 1.9917e-04 & 2.0477e-03 & 3.7849e-03 & 1.8536e-01 & 6.3829e-06 \\
\hline
$D_0=0$, $D_1=1$ & 3.1287e-05 & 1.0779e-02 & 5.2613e-07 & 1.2114e-03 & 2.6295e-03 & 5.1510e-03 & 2.2447e-01 & 4.1328e-06 \\
\hline
\hline
{\bf Loss: $55.1$ dB} & $p(D_0,D_1)$ & $e_{D_0D_1}$ &  $Y_{00,D_0D_1}^{\rm U}$ &  $Y_{11,D_0D_1}^{\rm U}$ &  $Y_{02,D_0D_1}^{\rm U}$ &  $Y_{20,D_0D_1}^{\rm U}$ & $e^{\rm ph}_{D_0D_1}$ & $\max_{\mu,\nu,\omega}R_{D_0D_1}$ \\
\hline
\hline
$D_0=1$, $D_1=0$ & 3.1489e-05 & 1.1575e-02 & 5.7419e-07 & 2.0888e-04 & 3.4223e-04 & 1.3150e-03 & 9.6353e-02 & 1.3762e-05 \\
\hline
$D_0=0$, $D_1=1$ & 3.1287e-05 & 1.0779e-02 & 4.8784e-07 & 1.1974e-03 & 1.5638e-04 & 2.6310e-03 & 1.1687e-01 & 1.1890e-05 \\
\hline
\end{tabular}
\caption{\label{tab:yields} Parameters $p(D_0,D_1)$, $e_{D_0D_1}$, $Y_{nm,D_0D_1}^{\rm U}$, $e^{\rm ph}_{D_0D_1}$ and $R_{D_0D_1}$, with $(D_0,D_1)\in\{(1,0),(0,1)\}$ and $(n,m)\in\{(0,0), (0,2), (2,0), (1,1)\}$, for the four different values of the overall system loss, 38.0 dB, 46.7 dB, 49.4 dB and 55.1 dB. For each value of the loss, we consider three different cases. The first one refers to the situation where we disregard the intensity fluctuations of the laser source, and the signal intensity $|\alpha|^2$ and the decoy intensities $\mu$, $\nu$ and $\omega$ take the mean values reported in Table I in the main text. The second (third) case refers to the situation where we consider intensity fluctuations and take as signal intensity and decoy intensities those, within the intervals reported in Table I in the main text, that minimise (maximise) the resulting secret key rate. These three cases are indicated in the table with the notation $R_{D_0D_1}$, $\min_{\mu,\nu,\omega}R_{D_0D_1}$ and $\max_{\mu,\nu,\omega}R_{D_0D_1}$, respectively.}
\end{table}

\begin{table}
	\begin{tabular}{|c|c|c|c|c|}
		\hline
		{\bf Loss: $38.0$ dB} & $p(D_0,D_1|0,0)$ & $p(D_0,D_1|0,1)$ & $p(D_0,D_1|1,0)$ & $p(D_0,D_1|1,1)$
		\\
		\hline
		\hline
		$D_0=1$, $D_1=0$ & 6.3640e-04 & 2.1371e-06 & 1.9464e-06 & 6.3243e-04 \\
		\hline
		$D_0=0$, $D_1=1$ & 1.7688e-06 & 6.2597e-04 & 6.2811e-04 & 2.7500e-06 \\
		\hline
		\hline
		{\bf Loss: $46.7$ dB} & $p(D_0,D_1|0,0)$ & $p(D_0,D_1|0,1)$ & $p(D_0,D_1|1,0)$ & $p(D_0,D_1|1,1)$
		\\
		\hline
		\hline
		$D_0=1$, $D_1=0$ & 2.3120e-04 & 1.2768e-06 & 1.4222e-06 & 2.3110e-04 \\
		\hline
		$D_0=0$, $D_1=1$ & 6.8021e-06 & 2.3240e-04 & 2.3070e-04 & 7.9559e-07 \\
		\hline
		\hline
		{\bf Loss: $49.4$ dB} & $p(D_0,D_1|0,0)$ & $p(D_0,D_1|0,1)$ & $p(D_0,D_1|1,0)$ & $p(D_0,D_1|1,1)$
		\\
		\hline
		\hline
		$D_0=1$, $D_1=0$ & 1.2588e-04 & 7.5870e-07 & 7.3259e-07 & 1.2525e-04 \\
		\hline
		$D_0=0$, $D_1=1$ & 7.2435e-07 & 1.2426e-04 & 1.2566e-04 & 6.9136e-07 \\
		\hline
		\hline
		{\bf Loss: $55.1$ dB} & $p(D_0,D_1|0,0)$ & $p(D_0,D_1|0,1)$ & $p(D_0,D_1|1,0)$ & $p(D_0,D_1|1,1)$
		\\
		\hline
		\hline
		$D_0=1$, $D_1=0$ & 6.2307e-05 & 6.9891e-07 & 7.5904e-07 & 6.2172e-05 \\
		\hline
		$D_0=0$, $D_1=1$ & 5.9439e-07 & 6.1977e-05 & 6.1837e-05 & 7.5494e-07 \\
		\hline
			
		\end{tabular}
	\caption{\label{tab:prob} Experimentally observed conditional probabilities $p(D_0,D_1|b_A,b_B)$ of the signals in the $X$ basis for the four different values of the overall system loss 38.0 dB, 46.7 dB, 49.4 dB and 55.1 dB, with $b_A,b_B \in \left\lbrace 0,1 \right\rbrace $ and $(D_0,D_1)\in \left\lbrace (1,0),(0,1)\right\rbrace $. }
\end{table}

\begin{table}
	\begin{tabular}{|c|c|c|c|c|c|c|c|c|c|}
		\hline
		{\bf Loss: $38.0$ dB} & $Q^{\mu_a\mu_b}_{D_0,D_1}$ & $Q^{\mu_a\nu_b}_{D_0,D_1}$ & $Q^{\mu_a\omega_b}_{D_0,D_1}$ & $Q^{\nu_a\mu_b}_{D_0,D_1}$ &  $Q^{\nu_a\nu_b}_{D_0,D_1}$ &  $Q^{\nu_a\omega_b}_{D_0,D_1}$ & $Q^{\omega_a\mu_b}_{D_0,D_1}$ & $Q^{\omega_a\nu_b}_{D_0,D_1}$ & $Q^{\omega_a\omega_b}_{D_0,D_1}$ \\
		\hline
		\hline
		$D_0=1$, $D_1=0$ & 1.1152e-03 & 6.3327e-04 & 5.7783e-04 & 6.2982e-04 & 1.1385e-04 & 5.9743e-05 & 5.7480e-04 & 5.7231e-05 & 1.8793e-06 \\
		\hline
		$D_0=0$, $D_1=1$ & 1.0701e-03 & 6.2515e-04 & 5.8166e-04 & 6.1052e-04 & 1.0991e-04 & 5.9878e-05 & 5.6856e-04 & 5.6217e-05 & 1.7243e-06 \\
		\hline
		\hline
		{\bf Loss: $46.7$ dB} & $Q^{\mu_a\mu_b}_{D_0,D_1}$ & $Q^{\mu_a\nu_b}_{D_0,D_1}$ & $Q^{\mu_a\omega_b}_{D_0,D_1}$ & $Q^{\nu_a\mu_b}_{D_0,D_1}$ &  $Q^{\nu_a\nu_b}_{D_0,D_1}$ &  $Q^{\nu_a\omega_b}_{D_0,D_1}$ & $Q^{\omega_a\mu_b}_{D_0,D_1}$ & $Q^{\omega_a\nu_b}_{D_0,D_1}$ & $Q^{\omega_a\omega_b}_{D_0,D_1}$ \\
		\hline
		\hline
		$D_0=1$, $D_1=0$ & 4.5761e-04 & 2.5417e-04 & 2.3487e-04 & 2.5735e-04 & 4.6880e-05 & 2.3929e-05 & 2.3521e-04 & 2.3785e-05 & 1.0445e-06 \\
		\hline
		$D_0=0$, $D_1=1$ & 4.5244e-04 & 2.5600e-04 & 2.4247e-04 & 2.5418e-04 & 4.6128e-05 & 2.4302e-05 & 2.3851e-04 & 2.3707e-05 & 7.5364e-07 \\
		\hline
		\hline
		{\bf Loss: $49.4$ dB} & $Q^{\mu_a\mu_b}_{D_0,D_1}$ & $Q^{\mu_a\nu_b}_{D_0,D_1}$ & $Q^{\mu_a\omega_b}_{D_0,D_1}$ & $Q^{\nu_a\mu_b}_{D_0,D_1}$ &  $Q^{\nu_a\nu_b}_{D_0,D_1}$ &  $Q^{\nu_a\omega_b}_{D_0,D_1}$ & $Q^{\omega_a\mu_b}_{D_0,D_1}$ & $Q^{\omega_a\nu_b}_{D_0,D_1}$ & $Q^{\omega_a\omega_b}_{D_0,D_1}$ \\
		\hline
		\hline
		$D_0=1$, $D_1=0$ & 6.8797e-05 & 4.9190e-05 & 3.5635e-05 & 4.8705e-05 & 2.8489e-05 & 1.4768e-05 & 3.5312e-05 & 1.4998e-05 & 7.0984e-07 \\
		\hline
		$D_0=0$, $D_1=1$ & 6.8811e-05 & 4.9155e-05 & 3.5687e-05 & 4.9062e-05 & 2.8450e-05 & 1.4764e-05 & 3.5386e-05 & 1.4924e-05 & 6.5292e-07 \\
		\hline
		\hline
		{\bf Loss: $55.1$ dB} & $Q^{\mu_a\mu_b}_{D_0,D_1}$ & $Q^{\mu_a\nu_b}_{D_0,D_1}$ & $Q^{\mu_a\omega_b}_{D_0,D_1}$ & $Q^{\nu_a\mu_b}_{D_0,D_1}$ &  $Q^{\nu_a\nu_b}_{D_0,D_1}$ &  $Q^{\nu_a\omega_b}_{D_0,D_1}$ & $Q^{\omega_a\mu_b}_{D_0,D_1}$ & $Q^{\omega_a\nu_b}_{D_0,D_1}$ & $Q^{\omega_a\omega_b}_{D_0,D_1}$ \\
		\hline
		\hline
		$D_0=1$, $D_1=0$ & 6.8058e-05 & 4.1643e-05 & 3.4735e-05 & 4.1441e-05 & 1.4647e-05 & 7.6051e-06 & 3.4506e-05 & 7.7697e-06 & 7.1022e-07 \\
		\hline
		$D_0=0$, $D_1=1$ & 6.7805e-05 & 4.1456e-05 & 3.4536e-05 & 4.1285e-05 & 1.4385e-05 & 7.3280e-06 & 3.4249e-05 & 7.5584e-06 & 6.2038e-07\\
		\hline
	\end{tabular}
	\caption{\label{tab:gains} Experimentally observed gains $Q^{ab}_{D_0,D_1}$ of the signals in the $Z$ basis for the four different values of the overall system loss, 38.0 dB, 46.7 dB, 49.4 dB and 55.1 dB, with $a,b\in \left\lbrace \mu,\nu,\omega \right\rbrace $ and $(D_0,D_1)\in \left\lbrace (1,0),(0,1)\right\rbrace $. Note that $a$ and $b$ refer to Alice's and Bob's intensity respectively.}
\end{table}

\begin{table}
\begin{tabular}{|c|c|c|c|c|}
\hline
\quad & $R_{10}+R_{01}$ & $\min_{\mu,\nu,\omega} R_{10}+R_{01}$ & $\max_{\mu,\nu,\omega} R_{10}+R_{01}$ & PLOB bound\\
\hline
\hline
{\bf Loss: $38.0$ dB} & 2.6484e-04 & 1.9917e-04 & 3.4765e-04 & 2.2867e-04 \\
\hline
{\bf Loss: $46.7$ dB} & 7.8389e-05 & 6.9058e-05 & 8.8458e-05 & 3.0845e-05 \\
\hline
{\bf Loss: $49.4$ dB} & 3.6306e-05 & 2.4061e-05 & 5.4130e-05 & 1.6564e-05 \\
\hline
{\bf Loss: $55.1$ dB} & 1.7542e-05 & 1.0516e-05 & 2.5652e-05 & 4.4584e-06 \\
\hline
\end{tabular}
\caption{\label{tab:Key} Secret key rate given by Eq.~(\ref{key_rate}) for the four different values of the overall system loss considered, 38.0 dB, 46.7 dB, 49.4 dB and 55.1 dB. For each value of the system loss, this Table includes three cases, {\it i.e.} the case where intensity fluctuations are disregarded and the worst and best case scenarios where intensity fluctuations are taken into account. These three cases are indicated in the table with the notation $R_{10}+R_{01}$, $\min_{\mu,\nu,\omega}R_{10}+R_{01}$ and $\max_{\mu,\nu,\omega}R_{10}+R_{01}$, respectively. They refer respectively, to the quantities  $R_{mean}$, $R_{min}$ and $R_{max}$ presented in the main text. Also, for comparison purposes, this table includes the PLOB bound~\cite{plob}. See the text for further details.}
\end{table}

%

%%%%%%%%%%%%%%%%%%%%%%%%%%%%%%%%%%%%%%%%%%%%%%%%%%%%%%%%%%%%%%%%%%%%
%\bibliographystyle{ieeetr}
%\bibliographystyle{unsrt}
% \bibliographystyle{apsrev}
%\bibliographystyle{apsrev4-1}
%\bibliography{FiniteHD}